\documentclass[aps,prl,reprint,showpacs,superscriptaddress,english,twocolumn,notitlepage,longbibliography,nofootinbib]{revtex4-1}
\usepackage{times}
\usepackage{graphicx}
\usepackage{float}
\usepackage{latexsym,amsmath,amssymb,bm,euscript}
\usepackage{color}
\usepackage{subfigure}
\usepackage{epstopdf}
\usepackage[colorlinks=true,linkcolor=blue,citecolor=blue,urlcolor=blue]{hyperref}
\usepackage{ulem}
\usepackage{cleveref}

\usepackage{tikz}
\usetikzlibrary{decorations.pathreplacing}

\usepackage{ytableau}

\definecolor{colorhhy}{rgb}{0.9, 0.17, 0.31}

\crefname{appendix}{Appendix}{Appendices}
\crefname{equation}{Eq.}{Eqs.}
\crefname{figure}{Fig.}{Figs.}
\crefname{table}{Table}{Tables}
\crefname{section}{Appendix}{Appendices}
\crefname{enumi}{Point}{Points}
\creflabelformat{appendix}{[#2#1#3]}
\crefmultiformat{figure}{Figs.~#2#1\xdef\mycreffirstarg{#1}#3}%
 { and~#2#1#3}{, #2#1#3}{ and~#2#1#3}

\newcommand{\be}[0]{\begin{equation}}
\newcommand{\ee}[0]{\end{equation}}

\def\ba#1\ea{\begin{align}#1\end{align}}

\newcommand{\bmat}[0]{\begin{bmatrix}}
\newcommand{\emat}[0]{\end{bmatrix}}

\def\kk{\mathbf{k}}

\newcommand{\citeSI}[1]{\cref{#1}}

\makeatletter
\let\oldcref\cref

\newcommand{\AddCrefMap}[2]{%
  \expandafter\def\csname crefmap@#1\endcsname{#2}%
}

\renewcommand{\cref}[1]{%
  \expandafter\ifx\csname crefmap@#1\endcsname\relax
    \oldcref{#1}%
  \else
    \csname crefmap@#1\endcsname
  \fi
}
\makeatother
\AddCrefMap{app:non_int_model}{Appendix~I~\cite{SM}}
\AddCrefMap{app:bosonization}{Appendix~II~\cite{SM}}
\AddCrefMap{app:SLL}{Appendix~III~\cite{SM}}
\AddCrefMap{app:FCI_phase_property}{Appendix~IV~\cite{SM}}
\AddCrefMap{app:FCI_interaction}{Appendix~V~\cite{SM}}
\AddCrefMap{app:RG}{Appendix~VI~\cite{SM}}
\AddCrefMap{app:nu_23}{Appendix~VII~\cite{SM}}

\begin{document}

\title{Quantum-Geometry-Induced Superconductivity near a Fractional Chern Insulator}

\author{Haoyu Hu}
\email{huhaoyu314@ustc.edu.cn}
\affiliation{Department of Physics, University of Science and Technology of China, Hefei, Anhui 230026, China}
\author{Lei Chen}
\affiliation{Department of Physics and Astronomy, Stony Brook University, Stony Brook, New York 11794, USA}
\begin{abstract}
Recent moiré experiments and numerical studies of interacting Chern bands have revealed fractional Chern insulators, charge-density-wave order, and superconductivity as proximate correlation-driven phases in topological systems.
How these phases compete or intertwine, and how quantum geometry shapes their interplay, remain open questions.
Here we present an analytic study of competing correlation-driven phases in a partially filled Chern band using a coupled-wire construction and bosonization.
The key ingredient is the coexistence of interaction channels that favor, respectively, a fractional Chern insulator (FCI) and a closely related anti-FCI (aFCI) state.
The aFCI channel is specific to lattice Chern bands and is enhanced by the quantum geometry of the Chern band.
We show that when both FCI and aFCI scattering channels are present, they cooperate and induce an effective coupling that drives a superconducting instability near the FCI phase.
The same mechanism can also favor a charge-density-wave phase, depending on microscopic parameters.
Using a perturbative renormalization-group analysis, we obtain the phase diagram and identify a superconducting regime adjacent to the FCI phase.
We further estimate the superconducting transition temperature and show that it is enhanced by quantum geometry. 
Our results establish quantum geometry as an organizing principle for the interplay among FCI, aFCI, and superconducting correlations.
\end{abstract}
\maketitle

{\it Introduction.---}
Partially filled Chern bands provide a setting where topology, interaction, and band geometry are inseparable. 
Repulsive interactions in such bands can stabilize fractional Chern insulators (FCIs)~\cite{PhysRevX.1.021014,Sheng2011,PARAMESWARAN2013816,PhysRevLett.106.236802,PhysRevLett.106.236803,PhysRevLett.106.236804}, which are lattice analogs of fractional quantum Hall states in the lowest Landau level (LLL). 
However, a Chern band can develop quantum geometry that deviates from the ideal LLL limit~\cite{PhysRevB.90.165139,PhysRevB.85.241308,Jackson2015,PhysRevLett.114.236802,PhysRevLett.127.246403}. 
This geometric structure can strongly affect correlation effects and distinguishes Chern-band systems from the LLL. 
Experimentally, fractional Chern insulators have been observed in moir\'e and graphene-based materials
~\cite{Cai2023,Zeng2023,Park2023,PhysRevX.13.031037,Ji2024,Redekop2024,Park2025,Lu2024,Kang2024,Wang2025,PhysRevX.15.011045}. 
In related material platforms, experimental evidence has also revealed other correlated phases, including superconductivity (SC) and charge-density-wave (CDW) order~\cite{Han2025,xu2026signaturesunconventionalsuperconductivitynear,Lu2025,PhysRevX.15.011045,sun2026twistangleevolutionvalleypolarizedfractional,sun2026twistangleevolutionvalleypolarizedfractional,Aronson2025}. 
Extensive efforts have also been made to understand the correlated phases in these materials~\cite{Bernevig2025,PhysRevB.107.L201109,PhysRevB.108.085117,ShavitOreg2024,PhysRevB.109.205121,PhysRevB.109.205122,PhysRevB.112.075109,PhysRevB.112.075110,PhysRevB.109.045147,PhysRevB.110.075109,PhysRevLett.133.206502,PhysRevLett.133.206503,PhysRevB.110.205124,PhysRevLett.133.206504,PhysRevB.110.115146,PhysRevX.14.041040,PhysRevLett.133.066601,PhysRevB.109.115116,PhysRevLett.132.236601,bernevig2025berrytrashcanmodelinteracting,zlqg-sj86,h22z-4hsj,qbg3-3yxv,sfm3-f1kp}. 
Recent numerical studies further suggest that SC and CDW phases can appear near FCI phases~\cite{doi:10.1073/pnas.2426680122,wang2025chiralsuperconductivitynearfractional,guerci2025fractionalizationchiraltopologicalsuperconductivity,Divic_2025,tm9q-w5y7,kuhlenkamp2025robust}. 
Several field-theoretical approaches and parton constructions have also been developed to understand how superconductivity may emerge from fractionalized excitations~\cite{kcm5-hx56,pichler2026microscopic,6bgj-bfdn,shi2025anyondelocalizationtransitionsdisordered,Divic_2025}. 
Together, these theoretical and experimental developments point to the possibility of a common microscopic setting from which FCI, CDW, and SC tendencies may emerge.
This possibility is especially intriguing because intrinsic superconductivity is not known to occur as a competing phase in the conventional LLL setting. 
One possible clue is the nontrivial quantum geometry of Chern bands. 
Despite this progress, a microscopic and analytical understanding of the interplay among FCI, SC, and CDW tendencies is still lacking. 
In this work, we provide such an analysis and show how quantum geometry can promote SC correlations near an FCI phase.

We study a partially filled Chern band in the anisotropic limit. 
In this limit, the two-dimensional system can be treated as an array of coupled wires and solved analytically using bosonization and the renormalization group (RG)~\cite{von1998bosonization,giamarchi2003quantum,cardy1996scaling}. 
This coupled-wire construction (CWC) has been widely used for LLL and Chern-band systems, where it captures the essential physics of fractional quantum Hall and fractional Chern insulating phases~\cite{PhysRevLett.88.036401,PhysRevB.99.035130,ShavitOreg2024,PhysRevB.89.085101,PhysRevB.90.201102,Meng2020}.  
Coupled-wire methods have also been applied to conventional correlated systems, such as Hubbard models, where they provide a useful way to study the competition and interplay among correlated phases, including superconductivity~\cite{PhysRevB.53.12133,PhysRevB.56.6569,PhysRevB.53.8521,PhysRevB.85.035104,RevModPhys.87.457}. 
We use this framework to analytically study the interplay among FCI, CDW, and superconducting tendencies, and to identify the role of quantum geometry~\cite{provost1980riemannian,RevModPhys.84.1419,Yu2025,PhysRevLett.131.240001}.

In our model, the quantum geometry is controlled by the ratio of the two inter-wire hopping amplitudes, $r=t_x'/t_x$, with increasing $r$ enhancing the quantum geometry.
Additionally, a finite $r$ leads to an additional anti-FCI (aFCI) scattering channel, which has also been identified in Ref.~\cite{ShavitOreg2024}. 
This aFCI channel is enhanced by quantum geometric effect and can destabilize the FCI phase.
We demonstrate that the cooperation between the aFCI and FCI channels generates an effective Josephson coupling between wires. 
This coupling produces a superconducting instability near the FCI phase. 
The same mechanism can also stabilize charge-density-wave order, depending on microscopic parameters. 
Using a RG analysis, we derive the resulting phase diagram and identify superconducting and charge-density-wave regimes near the FCI. 
Moreover, the superconducting transition temperature is generically enhanced by quantum geometry. 
Thus, our work provides an analytical study of the interplay among correlated phases, including FCI, superconductivity, and CDW order, and highlights the nontrivial role of quantum geometry. 

\begin{figure}
    \centering
    \includegraphics[width=1.0\linewidth]{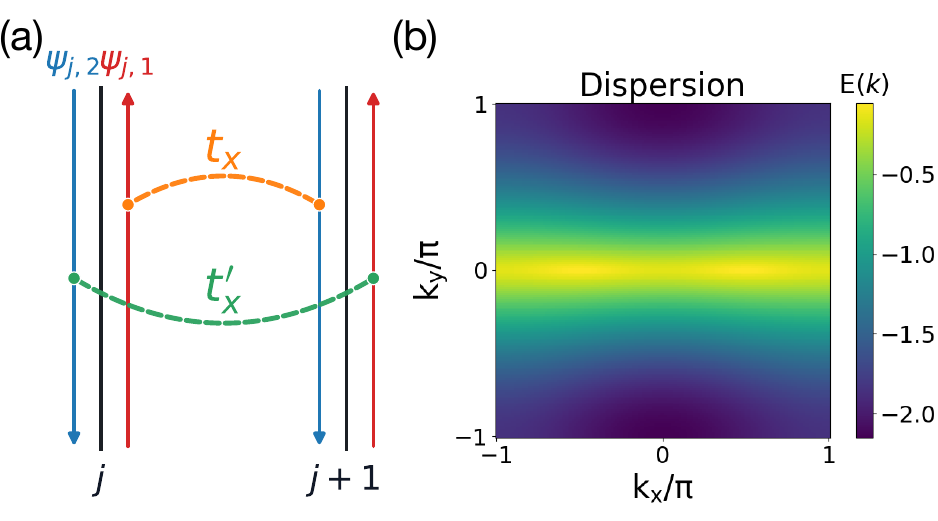}
    \caption{(a) Two nearest-neighbor inter-wire hopping processes with amplitudes $t_x$ and $t_x^{\prime}$. $\psi_{j,\alpha=1,2}$ denotes electron operators with opposite velocities on the $j$-th wire. 
    (b) Dispersion of the lowest band at $t_y=M=1$, $t_x=0.1, \mu=0$ and $r=0.5$. 
    }
    \label{fig:main_non_int}
\end{figure}

{\it Model and quantum geometry.---}
We take a minimal Chern-band model containing two $s$ orbitals related by inversion symmetry. 
The model is described by the following non-interacting Hamiltonian (see also \cref{app:non_int_model})
\ba
\label{eq:main_nonint_ham}
&H_0
=
\sum_{\kk,\alpha\gamma}
\bigg[\sum_\mu d_\mu(\kk)\tau_\mu-\mu\tau_0\bigg]_{\alpha\gamma}
c_{\kk,\alpha}^{\dag}c_{\kk,\gamma}\nonumber\\
&
d_y(\kk) = (-t_x +t_x')\sin(k_xa),\quad 
d_z(\kk) = t_y\sin(k_ya) \nonumber\\ 
& d_x (\kk) = M [1-\cos(k_ya)] + (t_x+t_x')\cos(k_xa). 
\ea
Here $\tau_\mu$ are Pauli matrices in orbital space, and $c_{\kk,\alpha}^\dag$ creates an electron with momentum $\kk$ and orbital index $\alpha$. 
We first consider the 1D limit with $t_y=M\ne0$ and $t_x=t_x'=0$, where we can view the system as a series of decoupled one-dimensional wire. 
On the $j$-th wire, at partial filling, the system has two Fermi points at $k_y = \pm k_F$. Expanding near these Fermi points gives two low-energy modes with opposite velocities, or chiralities, denoted by $\psi_{j,1}$ and $\psi_{j,2}$ (see \cref{fig:main_non_int} (a)).

We can then introduce weak inter-wire hopping between modes of opposite chirality. The two hopping amplitudes are $t_x$ and $t_x'$, as shown in \cref{fig:main_non_int}(a). At filling fraction $\nu=1$, the lowest band is fully filled, and these hopping processes stabilize a Chern insulator. The Chern number is $+1$ or $-1$, depending on whether $|t_x|>|t_x'|$ or $|t_x|<|t_x'|$. In this work, we focus on the anisotropic limit with a $C=1$ lowest band, where $t_y=M \gg |t_x| >|t_x'|$. A representative dispersion of the lowest band in this limit is shown in \cref{fig:main_non_int}(b).

This construction also makes clear how the lattice Chern band differs from the LLL in a magnetic field along the $z$ direction~\cite{Meng2020,ShavitOreg2024}.  
In the LLL-like limit, only the $t_x$ process is present~\cite{Meng2020}. 
In a Chern insulator, however, more hopping processes are allowed, and a finite $t_x'$ can be introduced. 
Therefore, we can increase $r= t_x'/t_x$ to drive the system away from the LLL limit. 
It is worth mentioning that the limit $t_x'=0$ was referred to as the optimal limit in Ref.~\cite{ShavitOreg2024}. This should not be confused with the ideal limit, since the trace condition is not satisfied even at $t_x'=0$.

\begin{figure}
    \centering
    \includegraphics[width=1.0\linewidth]{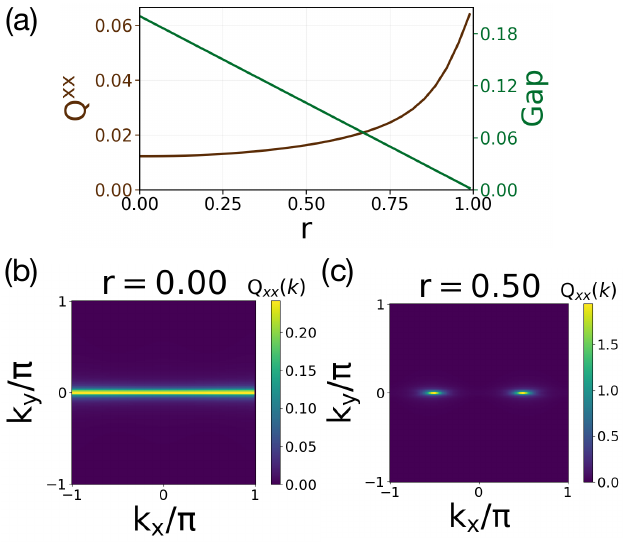}
    \caption{ (a) Integrated quantum geometry of the bottom band and the minimal indirect gap between the top and bottom bands as functions of $r$. Momentum dependence of $Q^{xx}(\kk)$ at $r=0.0$ (b) and $r=0.5$ (c), respectively. 
    We set $t_y=M=1$ and $t_x=0.1$.}
    \label{fig:main:QG}
\end{figure}

Moreover, we investigate the evolution of the quantum geometry. As $r$ increases, the integrated quantum metric $Q^{xx}$ along the $x$ direction grows, signaling a more delocalized wave function in the $x$ direction, as shown in \cref{fig:main:QG}(a). Analytically, $Q^{xx}$ takes the form
\ba 
Q^{xx} \sim \frac{\alpha_x}{2} \log\bigg( \frac{1}{\alpha_x} \frac{1+r^2}{|1-r^2|}
\bigg) ,\quad r = \frac{t_x'}{t_x}
\ea 
where $\alpha_x =\sqrt{8}\pi\sqrt{t_x^2+t_x'^2}/t_y $ is a small number measuring the anisotropy of the dispersion. 
As $r$ is increased from $0$, the quantum geometry is enhanced and then diverges logarithmically as $r\rightarrow 1$ ($t_x' \rightarrow t_x$). This behavior is consistent with the gap closing at $r=1$. 
Additionally, the momentum-space fluctuations of the quantum metric $Q^{xx}(\kk)$ increase with increasing $r$, as shown in \cref{fig:main:QG}(b) and (c). 
Thus, this toy model provides a minimal Chern-band setting with quantum geometry tunable by $r$, where stronger and more fluctuating quantum geometry is realized at larger $r$.

In the rest of the manuscript, we focus on the partially filled lowest $C=1$ band at filling fraction $\nu=1/3$. 
We start from the weakly coupled-wire limit, where the system is described by a sliding Luttinger liquid (SLL)~\cite{PhysRevLett.88.036401}. 
We then analyze how the interactions destabilize this SLL toward various competing phases, including a fractional Chern insulator and superconductivity.  
This coupled-wire construction also allows us to study the interplay among various phases analytically using bosonization. 
Within bosonization, the low-energy electron operator is written as $\psi_{j,\alpha} \sim K_{j,\alpha}e^{ -i \Phi_{j,\alpha}}$  (see \cref{app:bosonization} for details of the bosonization procedure). Here $K_{j,\alpha}$ is a Klein factor, and $\Phi_{j,\alpha}$ is the corresponding bosonic field. 
For later use, we also introduce $\theta_{j} = \frac{1}{2}(\Phi_{j,1} +\Phi_{j,2})$ and $\phi_{j} = \frac{1}{2}(\Phi_{j,1} -\Phi_{j,2})$.

\begin{figure}
    \centering
    \includegraphics[width=1.0\linewidth]{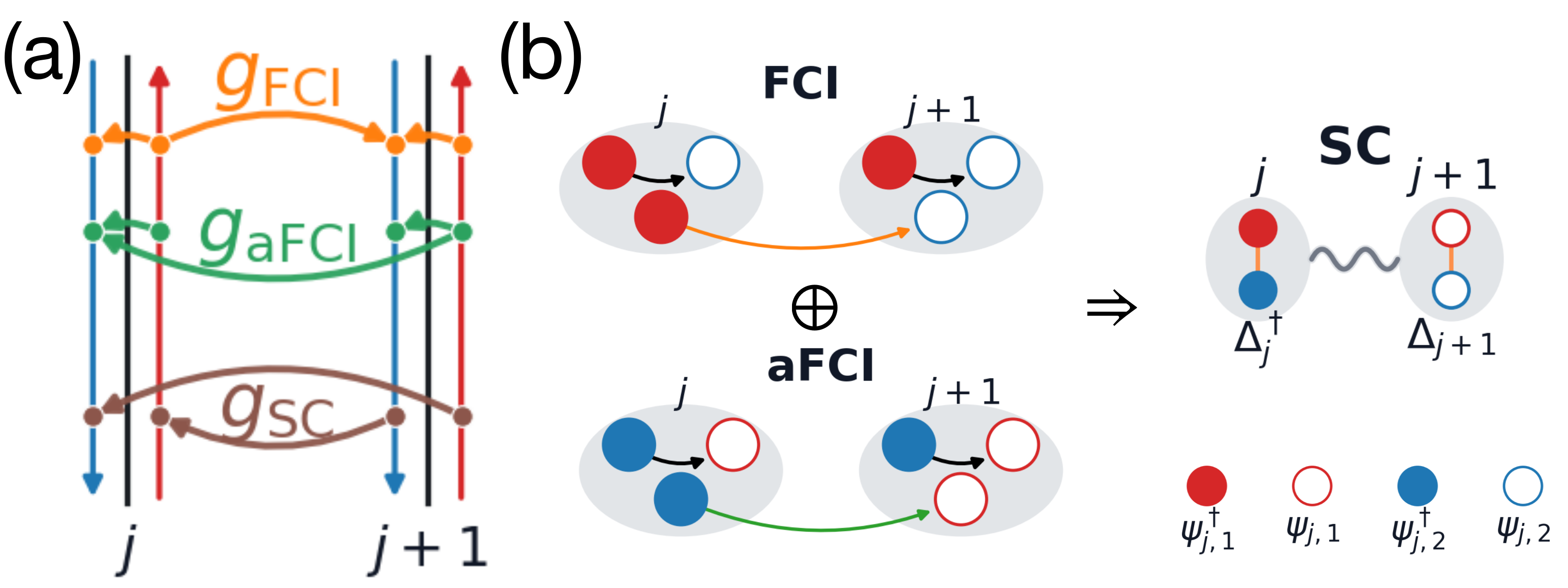}
    \caption{(a) Scattering processes associated with FCI, aFCI, and superconducting channels.  (b) Combining $e^{i\Theta_j^{FCI}}$ with $e^{-i\Theta_j^{aFCI}}$ leads to a Josephson coupling $\Delta_j^{\dag}\Delta_{j+1}$ between neighboring wires.}
    \label{fig:sc_process}
\end{figure}

{\it Fractional Chern insulator.---}
At $\nu=1/3$, the FCI is stabilized by the following effective correlated hopping process, as illustrated schematically in \cref{fig:sc_process}(a), 
\ba 
\label{eq:main:FCI_scattering}
\text{FCI}:\quad &\sum_j[ O_j(r) \psi_{j,1}^\dag(r+a) ] [O_{j+1}(r) \psi_{j+1,2}(r+a)], 
\ea 
where the particle-hole operator is defined as $O_j(r)=\psi_{j,1}^\dag(r)\psi_{j,2}(r)$. The displacement $a$ denotes a small real-space separation. 
Besides FCI scattering channel, there is another closely related scattering process, denoted aFCI, which takes the form (see also \cref{fig:sc_process}(a)) 
\ba 
\label{eq:main:aFCI_scattering}
\text{aFCI}:\quad &\sum_j[ O_j(r) \psi_{j,2}(r+a) ] [O_{j+1}(r) \psi_{j+1,1}^\dag(r+a)].
\ea

Microscopically, the FCI interaction is induced by the combination of on-site inter-orbital repulsion with strength $U$ and inter-wire hopping characterized by $t_x$. 
The aFCI interaction is generated by the analogous process with the inter-wire hopping replaced by the $t_x'$ process. 
The coupling strengths obtained from perturbation theory read (see \cref{app:FCI_interaction} for details of the derivation)
\ba 
\label{eq:main:def_g_FCI_aFCI}
&g_{FCI} \propto \frac{U^2Q^{yy}(k_F)}{E_H^2}t_x
,\quad g_{aFCI} \propto g_{FCI} r, 
&
\ea  
where $E_H$ denotes the energy of the high-energy electrons. $Q^{yy}(k_F)$ is the quantum geometry along the $y$ direction at the Fermi point in the 1D limit.
Enhanced quantum geometry along $y$ therefore enhances the overall coupling strength. 
Here, we are mostly interested in the relative strength between FCI and aFCI scattering. This relative strength controls the stability of the FCI phase and plays an important role in stabilizing the superconducting phase discussed below. As shown in \cref{eq:main:def_g_FCI_aFCI}, the relative coupling strength between FCI and aFCI is $r$, which is also directly related to the quantum geometry $Q^{xx}$. 
Intuitively, when the electronic wave functions become more delocalized along the $x$ direction, as characterized by the enhancement of $Q^{xx}$, the additional correlated hopping process additional $x$ direction, represented by the aFCI channel, emerges and becomes more important.

Finally, within the bosonization framework, the interactions given in \cref{eq:main:FCI_scattering,eq:main:aFCI_scattering} correspond to the interaction vertices 
\ba 
g_{FCI}e^{i\Theta_{j}^{FCI}(r)},\quad g_{aFCI}e^{i\Theta_{j}^{aFCI}(r)}\, ,
\ea 
where the corresponding bosonic fields are 
$\Theta_j^{FCI/aFCI}(r)= \pm [\theta_j(r) -\theta_{j+1}(r)] + 3(\phi_j(r)+\phi_{j+1}(r))$.

{\it Superconducting (and charge-density-wave) instability.---}
We now show that superconducting and charge-density-wave correlations are naturally generated when the FCI and aFCI scattering processes are both present. This follows from the operator product expansion (OPE)~\cite{cardy1996scaling},
\ba 
\label{eq:main:OPE}
&: e^{i\Theta_j^{FCI}(R-\frac{r}{2})}:
:e^{-i\Theta_j^{aFCI}(R+\frac{r}{2})}: 
\sim 
\frac{e^{i\Theta_j^{SC}(R)}}{|r|^{\Delta^{FCI}+\Delta^{aFCI}-\Delta^{SC}}} \nonumber\\ 
&: e^{i\Theta_j^{FCI}(R-\frac{r}{2})}:
:e^{i\Theta_j^{aFCI}(R+\frac{r}{2})}: 
\sim 
\frac{e^{i\Theta_j^{CDW}(R)} }{|r|^{\Delta^{FCI}+\Delta^{aFCI}-\Delta^{CDW}}}
\ea 
where $::$ denotes normal ordering and $\Delta^\lambda$ denotes the scaling dimension of $\Theta_j^\lambda$ with $\lambda \in \{FCI,aFCI,SC,CDW\}$~\cite{cardy1996scaling} (see \cref{app:RG} for the derivation). 
From \cref{eq:main:OPE}, we observe that the fusion between $e^{i\Theta^{FCI}}$ and $e^{\pm i \Theta^{aFCI}}$ naturally leads to interaction vertices that we denote as SC and CDW, with the corresponding bosonic fields defined as 
\ba 
&\Theta_j^{SC}(r) = \Theta_j^{FCI}(r) - \Theta_j^{aFCI}(r), 
\nonumber\\ 
&\Theta_j^{CDW}(r) = \Theta_j^{FCI}(r) + \Theta_j^{aFCI}(r) .
\ea 
To identify the physical content of $\Theta_j^{SC}$ and $\Theta_j^{CDW}$, we transform back to the electronic fields, where we find 
\ba 
&e^{i\Theta_j^{SC}(r)} \sim \Delta_j^\dag(r)  \Delta_{j+1}(r),\quad \Delta_j(r) = \psi_{j,1}(r) \psi_{j,2}(r) \nonumber\\
&
e^{i\Theta_j^{CDW}(r)} \sim [O_j(r)]^3  [O_{j+1}(r)]^3 ,\quad 
O_j(r)=\psi_{j,1}^\dag(r)\psi_{j,2}(r)
\ea 
Thus $e^{i\Theta_j^{SC}(r)}$ describes a Josephson coupling between pairing fields on neighboring wires, which promotes coherent pairing correlations across the wire array and stabilizes an SC phase. By contrast, $e^{i\Theta_j^{CDW}(r)}$ locks particle-hole correlations between wires and stabilizes a CDW phase. 

Qualitatively, the emergence of the SC channel is illustrated in \cref{fig:sc_process} (b). 
We start by combining the $e^{i\Theta^{FCI}_j}$ process with the conjugate aFCI process $e^{-i\Theta^{aFCI}_j}$. 
After canceling electron and hole operators within the same wire and chirality, the remaining operator contains a pair of electron operators on wire $j$ and a pair of hole operators on wire $j+1$, corresponding to a Josephson coupling between the two wires.  
Intuitively, the OPE shows that correlations in the FCI and aFCI channels induce correlations in the SC and CDW channels. 

{\it Perturbative RG.---}
After establishing the connections among FCI, aFCI, SC, and CDW instabilities, we perform a perturbative RG calculation to further identify their interplay. 
We focus on a two-wire limit, where the system consists of two wires with $j=1,2$. This limit is analytically tractable and is also sufficient to capture the competition and intertwining among these channels. 
Within the two-wire limit, the free-boson part is described by
\ba 
\label{eq:main:two_wire_SLL}
H_{SLL} 
= &\int \frac{dr}{2\pi}\bigg\{\sum_{i \in \{e,o\} }u_i
\bigg[ 
K_i [\partial_r \theta_i(r)]^2 
+ \frac{1}{K_i} [\partial_r \phi_i(r)]^2 
\bigg] \nonumber\\ 
&
+v_1 \partial_r \theta_o(r) 
\partial_r \phi_e(r) 
+ v_2 \partial_r \theta_e(r) 
\partial_r \phi_o(r) 
\bigg\} 
\ea 
where $\theta_{e/o} =\frac{1}{2}(\theta_{1} \pm \theta_2)$ and $\phi_{e/o}=\frac{1}{2}(\phi_{1} \pm \phi_2)$. 
$K_{i}$ are the corresponding Luttinger parameters, $u_{i}$ is the velocity, $v_1$ and $v_2$ are symmetry-allowed couplings in the time-reversal-breaking system (see \citeSI{app:RG} for the corresponding values derived from the microscopic model).
We consider interactions $g_\lambda e^{i\Theta_{j}^\lambda}+\text{h.c.}$ with $\lambda \in \{FCI,aFCI,SC,CDW\}$ as perturbations. 
Here $g_\lambda$ is the corresponding coupling constant. 
To facilitate the RG calculation, we also introduce dimensionless couplings $y_{\lambda} \propto g_\lambda$. The RG flows of the coupling constants are
\ba 
\label{eq:main:RG_flow}
&\partial_l y_{FCI} = (2-\Delta^{FCI})y_{FCI}
-
(y_{aFCI}y_{CDW} + y_{SC}y_{aFCI} ) \nonumber\\
&\partial_l y_{aFCI} = (2-\Delta^{aFCI})y_{aFCI}
-
(y_{FCI}y_{CDW} + y_{SC}y_{FCI} ) \nonumber\\
&\partial_l y_{SC} = (2-\Delta^{SC})y_{SC}
-
y_{FCI} y_{aFCI} \nonumber\\
&\partial_l y_{CDW} = (2-\Delta^{CDW})y_{CDW}
-
y_{FCI} y_{aFCI}
\ea 
where $l$ denotes the RG step. 
The full RG equations and their derivation are provided in \cref{app:RG}.

Equation~\eqref{eq:main:RG_flow} shows that SC and CDW couplings are generated under RG once both FCI and aFCI scatterings are present with $y_{FCI}y_{aFCI}\ne 0$. 
Importantly, a finite SC coupling can emerge under the RG flow even when the microscopic Hamiltonian contains no interaction channel that directly favors superconductivity.
The scaling dimensions $\Delta^{SC}=2/K_o$ and $\Delta^{CDW}=18K_e$ control whether the generated SC or CDW coupling becomes the leading instability. 

\begin{figure}
    \centering
    \includegraphics[width=1.0\linewidth]{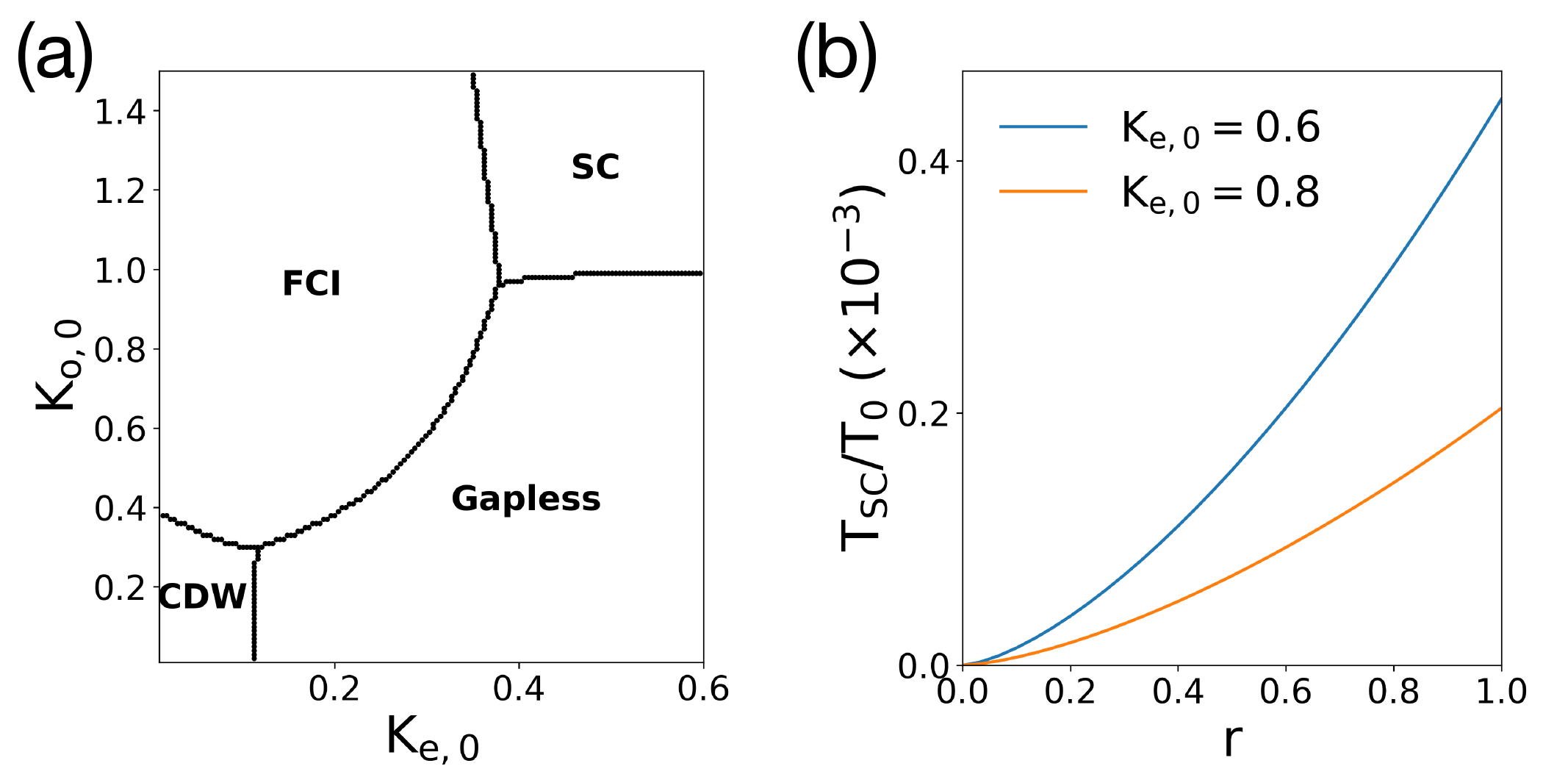}
    \caption{RG phase diagram and SC transition temperature. (a) Leading instability obtained from the perturbative RG flow. The microscopic coupling constant before the RG procedure is taken to be $K_e|_{l=0}=K_{e,0}$, $K_o|_{l=0}=K_{o,0}$, $y_{FCI}|_{l=0}=0.1$, $y_{aFCI}|_{l=0}=0.05$, and $y_{SC}|_{l=0}=y_{CDW}|_{l=0}=0$, corrsponging to $r=0.5$. (b) Estimated SC transition temperature $T_{SC}/T_0$ as a function of $r$, evaluated in the SC regime with $K_{o,0}=1.5$ and $K_{e,0}=0.6,0.8$. The increase of $T_{SC}$ tracks the strengthening of quantum geometry. }
    \label{fig:main_phase_diagram}
\end{figure}

We solve the full set of RG equations in the absence of microscopic couplings in the SC and CDW channels, while keeping finite bare couplings in the FCI and aFCI channels. The resulting phase diagram is shown in \cref{fig:main_phase_diagram}(a), where the ground state is determined by which dimensionless coupling $y_\lambda$ first reaches $1$ during the RG flow. 
A gapless region remains when none of the scattering processes becomes relevant. 
Both SC and CDW instabilities naturally appear near the FCI region. 
In addition, we find that a stronger single-particle dispersion gives rise to larger $K_{o,0}$ and $K_{e,0}$. This tends to favor SC over CDW once the FCI phase becomes unstable.

Finally, we note that the overall structure of the phase diagram remains qualitatively similar as long as a finite $r$ is included, which allows for a finite interaction strength in the aFCI channel. However, the superconducting transition temperature is sensitive to $r$, increasing as the quantum-geometry effect becomes more pronounced. 
Within the RG flow, the SC transition temperature can be estimated from $T_{SC}=T_0e^{-l^*_{SC}}$. Here $T_0$ is the ultraviolet energy scale. $l^*_{SC}$ is the RG step at which the dimensionless coupling $y_{SC}$ first reaches $1$, indicating that SC correlations become relevant. 
In \cref{fig:main_phase_diagram}(b), we illustrate how the $T_{SC}$ increases as $r$ is increased. 
We note that, enhancing quantum geometry by increasing $r$ enhances the $y_{aFCI}$ coupling. As a consequence, the SC tendency is strengthened by the FCI/aFCI product term in the RG flow (\cref{eq:main:RG_flow}). 
 
{\it Filling $\nu=2/3$.---}
We finally mention that a similar mechanism can also drive an SC instability at filling $\nu=2/3$. At this filling, the bosonized fields stabilizing the FCI phase can be obtained by particle-hole conjugation~\cite{PhysRevB.99.035130}.
We consider both the FCI interaction and the related aFCI interaction, characterized by the bosonic fields  $\Theta_j^{\overline{FCI}/\overline{aFCI}} =
 4\phi_j +\phi_{j-1} +\phi_{j+1}\pm  [\theta_{j-1}-\theta_{j+1}]
$ (see \cref{app:nu_23} and Ref.~\cite{PhysRevB.99.035130} for further discussion of particle-hole conjugation). 
Combining the two scattering processes again generates SC and CDW correlations, with
$
\Theta_j^{\overline{SC}/\overline{CDW}}(r) = \Theta_j^{\overline{FCI}}(r) \mp  \Theta_j^{\overline{aFCI}}(r) 
$ 
whose fermionic representations are
\ba 
&e^{i\Theta_j^{\overline{SC}}(r)} \sim \Delta_{j-1}^\dag(r) \Delta_{j+1}(r) 
\nonumber\\
&e^{i\Theta_j^{\overline{CDW}}(r)} \sim  O_{j-1}(r) [O_{j}(r)]^2 O_{j+1}(r) 
\ea 
Thus, SC and CDW correlations can again be induced by the cooperation of FCI and aFCI scattering. However, the nature of the resulting phases is different from the $\nu=1/3$ case. This is because the FCI-favoring term at $\nu=2/3$ involves three wires. Consequently, the resulting Josephson coupling and CDW locking involve next-nearest-neighbor wires.

{\it Summary and discussion.---}
In summary, we have provided an analytical study of competing phases in a partially filled interacting Chern band.
In our model, the quantum geometry is controlled by the parameter $r$, and increasing $r$ enhances the quantum geometry. 
As the quantum geometry is enhanced, a competing aFCI scattering process emerges and becomes stronger. 
The cooperation between the aFCI and FCI channels naturally produces correlations in the SC and CDW channels.
Using a perturbative RG analysis, we derive the phase diagram of the model and find SC and CDW phases near the FCI phase. 
We further demonstrate that the estimated SC transition temperature is enhanced by the quantum-geometry effect. 
We show that SC and CDW correlations can also be induced by the cooperation of FCI and aFCI scattering at filling fraction $\nu=2/3$. 
Therefore, our work provides an analytical study of the interplay among FCI, SC, and CDW phases, and identifies the crucial role of quantum geometry in organizing these instabilities. 

\paragraph*{Acknowledgments --- } This work was performed in part at the Aspen Center for Physics, which is supported by National Science Foundation grant PHY-2210452 and by a grant from the Simons Foundation (1161654, Troyer). 

\begingroup
  \renewcommand{\addcontentsline}[3]{}
  \begin{acknowledgments}
  \end{acknowledgments}
\endgroup

\bibliography{ref}

\onecolumngrid
\newpage 
\begin{center}
\textbf{Supplementary Materials}\\ 
\end{center}

\setcounter{secnumdepth}{3}
\setcounter{section}{0}
\setcounter{subsection}{0}
\setcounter{subsubsection}{0}
\setcounter{figure}{0}
\setcounter{table}{0}
\setcounter{equation}{0}

\renewcommand{\thefigure}{S\arabic{figure}}
\renewcommand{\theHfigure}{supp.figure.\arabic{figure}}

\renewcommand{\thetable}{S\arabic{table}}
\renewcommand{\theHtable}{supp.table.\arabic{table}}
\renewcommand{\thesection}{\Roman{section}}
\renewcommand{\theHsection}{supp.section.\arabic{section}}
\renewcommand{\thesubsection}{\thesection.\arabic{subsection}}
\renewcommand{\theHsubsection}{supp.subsection.\arabic{section}.\arabic{subsection}}
\renewcommand{\thesubsubsection}{\thesubsection.\arabic{subsubsection}}
\renewcommand{\theHsubsubsection}{supp.subsubsection.\arabic{section}.\arabic{subsection}.\arabic{subsubsection}}

\renewcommand{\theequation}{S\arabic{equation}}
\renewcommand{\theHequation}{supp.equation.\arabic{equation}}

\makeatletter
\let\app@oldsection\section
\let\app@oldsubsection\subsection
\renewcommand{\section}{\@ifstar{\app@oldsection*}{\app@section}}
\newcommand{\app@section}[1]{%
  \refstepcounter{section}%
  \setcounter{subsection}{0}%
  \app@oldsection*{\texorpdfstring{\thesection\quad #1}{\thesection\space #1}}%
}
\renewcommand{\subsection}{\@ifstar{\app@oldsubsection*}{\app@subsection}}
\newcommand{\app@subsection}[1]{%
  \refstepcounter{subsection}%
  \app@oldsubsection*{\texorpdfstring{\thesubsection\quad #1}{\thesubsection\space #1}}%
}
\makeatother

\tableofcontents

\clearpage





\section{Non-interacting Hamiltonian}
\label{app:non_int_model}

We consider a toy model with $s$ and $p$ orbitals located at the $1a$ position and labeled by $c_{\kk,s/p}$, respectively. We take the layer group to be $p\bar{1}$, with translational and inversion symmetries. 
It is useful to recombine the $s$ and $p$ orbitals into two $s$ orbitals,
 \ba 
c_{\kk, 1/2} = \frac{1}{\sqrt{2}}(c_{\kk,s} \pm c_{\kk,p})
 \ea 
which satisfy the following symmetry property
 \ba 
 I c_{\kk,\alpha }I^{-1} = c_{-\kk,3-\alpha }
 \ea 
where $I$ denotes the inversion symmetry.

We further consider the anisotropic limit, in which the system develops strong dispersion along the $y$ direction. This allows us to treat the system within the wire construction. 
The minimal symmetry-allowed non-interacting Hamiltonian can be written as 
\ba 
\label{eq:non_int_ham}
H_0 = 
\sum_{\kk,\alpha\gamma}  \bigg[ t_y \sin(k_ya) \tau_z + [M (1-\cos(k_ya))+ (t_x+t_{x'}) \cos(k_xa)] \tau_x 
+ (-t_x+t_{x'}) \sin(k_xa) \tau_y
-\mu \tau_0 \bigg]_{\alpha ,\gamma }c_{\kk,\alpha}^\dag c_{\kk,\gamma}
\ea 
where $\tau_{x,y,z}$ are Pauli matrices in the orbital space and $\mu$ is the chemical potential. 
The dispersion is 
\ba 
\label{eq:nonint_disp}
E_{\kk,\pm} = \pm \sqrt{ [t_y\sin(k_ya)]^2 
+ [M (1-\cos(k_ya))  
+  (t_x+t_x') \cos(k_xa) 
]^2  + [ (t_x-t_x') \sin(k_xa)]^2
} -\mu 
\ea 

\begin{figure}
    \centering
    \includegraphics[width=0.9\linewidth]{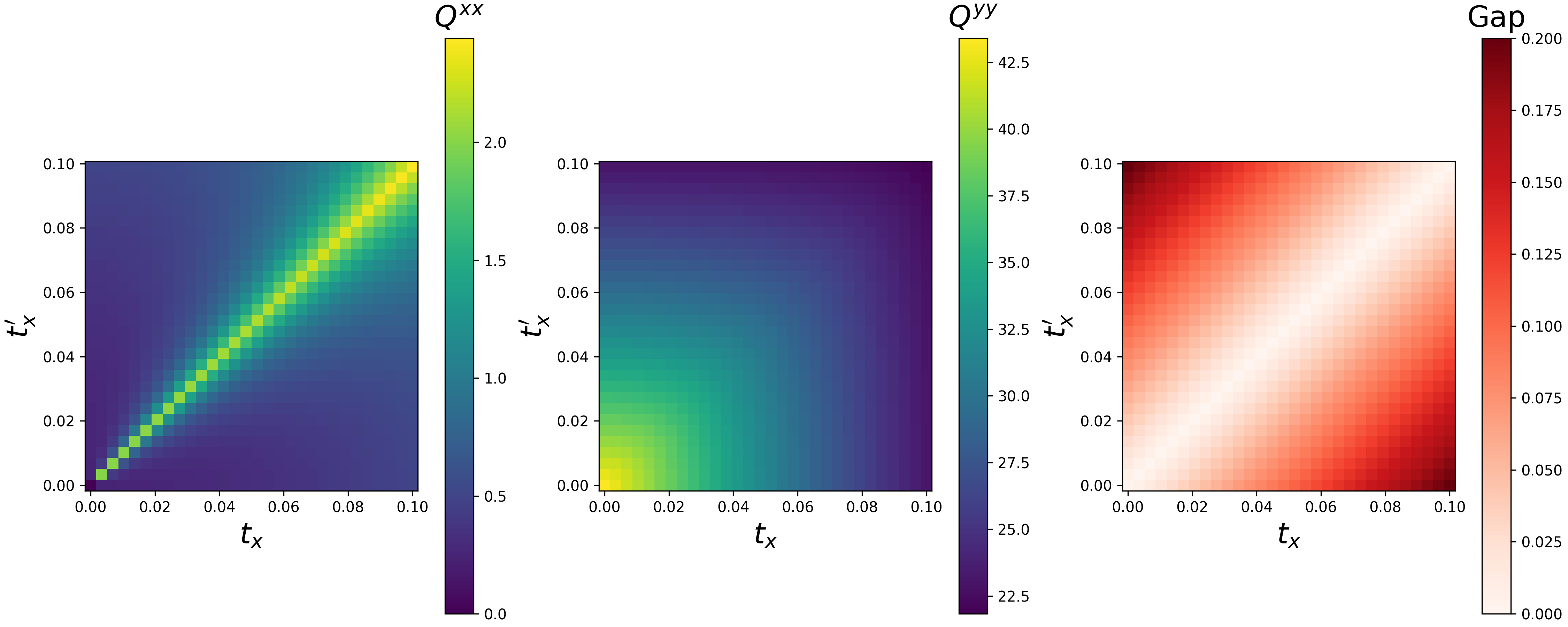}
    \caption{Quantum geometry (\cref{eq:def_QG}) and the gap between two bands as a function of $t_x, t_x'$ with $t_y = M = 1$. }
    \label{fig:QG_gap_scan}
\end{figure}

We consider the parameter regime where $|t_y|,|M| \gg  |t_x|,|t_{x'}|$. 
In this anisotropic limit, the Chern number of the lowest band reads 
\ba 
\begin{cases}
    C= 1 & |t_x| > |t_{x}'| \\
    C= -1 & |t_x| <|t_{x}'| 
\end{cases}
\ea 
We also introduce the quantum geometry (Fubini–Study metric) of the lowest band, which characterizes the real-space delocalization of the wave function
\ba 
\label{eq:def_QG}
Q^{\mu\mu} = \frac{1}{4\pi^2} \int dk_xdk_y Q^{\mu\mu}(\kk),\quad Q^{\mu\mu}(\kk) 
= \langle \partial_{k_\mu}u_\kk| \partial_{k_\mu} u_\kk\rangle 
- \langle \partial_{k_\mu} u_\kk|u_\kk\rangle \langle u_\kk |\partial_{k_\mu} u_\kk\rangle 
\ea 
with $|u_\kk\rangle$ the corresponding Bloch wave function.
The quantum geometry and the minimal gap between the two bands are controlled by $t_x/t_x'$, as illustrated in \cref{fig:QG_gap_scan}. Several general remarks are in order
\begin{itemize}
    \item We are primarily interested in the quantum geometry along the $x$ direction, which characterizes the spatial delocalization of electrons in that direction.
    \item The quantum geometry is enhanced when the gap reaches its minimum. Thus strong quantum geometry appears near the gap-closing point with $|t_x| = |t_x'|$. 
\end{itemize}

We also provide some analytical insight into the behavior of $Q^{xx}(\kk)$. 
The dominant contribution to the quantum geometry comes from $k_y=0$, where the gap reaches its minimum. Near $k_y=0$, we can approximate $t_y\sin(k_ya) \approx t_y k_y a$ and $M(1-\cos(k_ya)) \approx 0 $. This leads to 
\ba 
Q^{xx}(k_x,k_y)=a^2 
\frac{( t_x^2-t_x'^2)^2 + m_{k_y}^2 (t_x^2 +t_x'^2)-2m_{k_y}^2 t_x t_x'\cos(2ak_x)}{ 
4 \bigg[m_{k_y}^2 + t_x^2 + t_x'^2 + 2t_x t_x'\cos(2 ak_x)\bigg]^2  
},\quad m_{k_y} \approx t_y k_y a
\ea 
Integrating over $k_x$ and $k_y$ gives 
\ba 
Q^{xx}= & \frac{1}{4\pi^2}\int dk_y \int dk_x Q^{xx}(k_x,k_y) \approx \frac{1}{4\pi^2}
\int_{-\pi/a}^{\pi/a} dk_y \frac{a}{2}\frac{\pi (t_x^2+t_x'^2) }{\sqrt{-4t_x^2t_x'^2 + (m_{k_y}^2 +t_x^2+t_x'^2)^2}} \nonumber\\
\approx & 
 \frac{1}{4\pi^2}
\int^{\pi/a}_{-\pi/a} dk_y \frac{a}{2}\frac{\pi (t_x^2+t_x'^2) }{\sqrt{ 
(t_x^2-t_x'^2)^2 + 2m_{k_y}^2 ( t_x^2 + t_x'^2) } }  \nonumber\\
\approx & 
\frac{1}{4\pi} \frac{\sqrt{t_x^2+t_x'^2}}{\sqrt{2}t_y}\text{arcsinh}
\bigg( 
\frac{\sqrt{2}\pi t_y\sqrt{t_x^2+t_x'^2} }{|t_x^2-t_x'^2|} 
\bigg) 
\ea 
To simplify the notation, we let 
\ba 
r = \frac{t_x'}{t_x} 
\quad \alpha_x = \frac{\sqrt{t_x^2+t_x'^2}}{t_y} \frac{1}{2\sqrt{2}\pi}
\ea 
In the anisotropic limit with small $\alpha_x$, we find 
\ba 
Q^{xx} \approx  \frac{\alpha_x}{2}\text{arcsinh}
\bigg( 
\frac{ {1+r^2} }{2\alpha_x (1-r^2)}  
\bigg) 
\approx \frac{\alpha_x}{2}
\log(\frac{1 }{\alpha_x}\frac{1+r^2}{1-r^2})
\ea 
showing that the quantum geometry diverges logarithmically as $r\rightarrow 1^-$. This is expected, since the gap closes at $r= 1$.  



Throughout this manuscript, we focus on the anisotropic limit in which the lowest band forms a $C=1$ Chern band with 
\ba 
&|t_y|,|M| \gg |t_x|, |t_x'| \nonumber\\
&|t_x| > |t_x'|  
\ea 
We then study the interacting physics of this partially filled lowest band.

\section{Coupled wire construction and bosonization}
\label{app:bosonization}

We focus on the limit
\ba 
|t_y|,|M | \gg  |t_x| ,|t_x'| 
\ea 
where the system develops strong dispersion along the $y$ direction and weak coupling along the $x$ direction. It is then useful to view the system as a series of weakly coupled one-dimensional wires and treat it using bosonization.

On the $j$-th wire, the corresponding electron operators are 
\ba 
c_{j,\alpha,k_y}^\dag = \frac{1}{\sqrt{N_x}}\sum_{k_x} c_{\alpha,(k_x,k_y)}e^{ik_xja }
\ea 
where $N_x$ is the number of sites in the $x$ direction. 
The Hamiltonian of each wire in the decoupled limit ($t_x=t_x'=0$) reads 
\ba 
H_{0,j} = \sum_{k_y,\alpha\gamma} 
  \bigg[ t_y \sin(k_ya) \tau_z + [M (1-\cos(k_ya))] \tau_x 
-\mu \tau_0 \bigg]_{\alpha \gamma }c_{j,\alpha,k_y}^\dag c_{j,\gamma,k_y}
\ea 
The dispersions are 
\ba 
E_{\kk,\pm}= -\mu \pm \sqrt{ [M (1-\cos(k_ya))]^2 + [t_y\sin(k_ya)]^2
} 
\ea 

The corresponding band-basis operator for the bottom band $(\gamma_{j,k_y})$ is
\ba 
\label{eq:band_basis_operator}
{\gamma}_{j,k_y}^\dag = 
\frac{ \bigg[ d_z(k_y) + \sqrt{[d_x(k_y)]^2 + [d_z(k_y)]^2 }\bigg] c_{j,2,k_y}^\dag 
- d_x(k_y) c_{j,1, k_y}^\dag 
}{\sqrt{ 2 \sqrt{[ d_x(k)]^2 + [d_z(k)]^2 } 
\bigg[ 
\sqrt{[ d_x(k)]^2 + [d_z(k)]^2 }
+d_z(k) 
\bigg] 
}}
\ea 
with 
\ba 
d_z(k_y) = t_y \sin(k_ya),\quad d_x(k_y) 
= M(1-\cos(k_ya)) 
\ea

\begin{figure}
    \centering
    \includegraphics[width=0.6\linewidth]{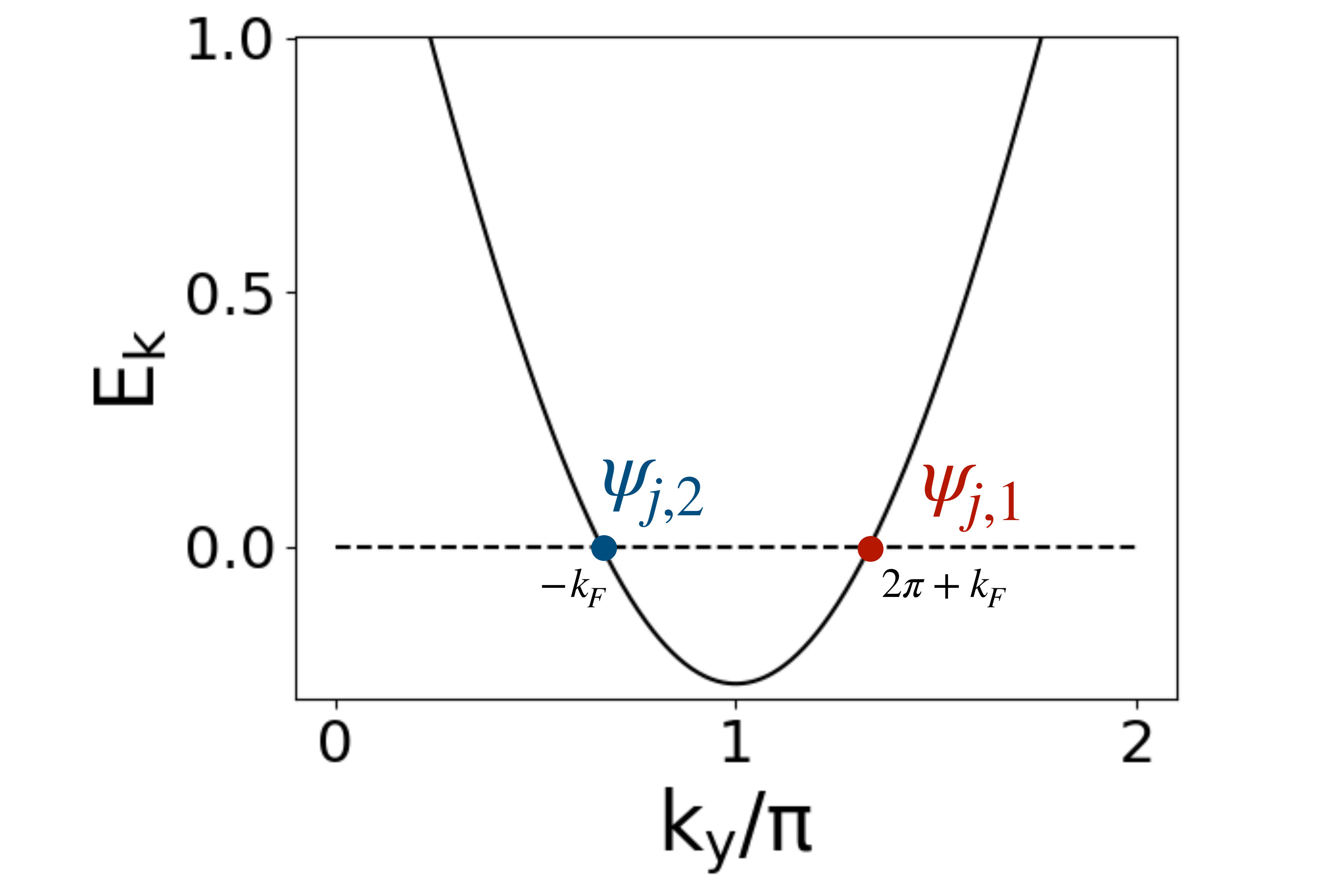}
    \caption{Illustration of the one-dimensional band structure for $t_y=M=1$. The Fermi energy corresponds to filling fraction $\nu=1/3$.}
    \label{fig:1d_disp}
\end{figure}

We consider filling factor $\nu$ in the bottom band with Chern number $+1$. The corresponding Fermi momentum is 
\ba 
\pm k_F = \mp  (1-\nu) \pi /a 
\ea 
We retain only the low-energy degrees of freedom near $k_F$, for which 
\ba 
\label{eq:H_j0}
H_{0,j} 
\approx 
\sum_{ p_y, \alpha }
v (-1)^{\alpha+1} \gamma_{j,(-1)^{\alpha+1}k_F + p_y}^\dag \gamma_{j,(-1)^{\alpha+1}k_F+p_y} 
\ea 
where $v$ is the corresponding Fermi velocity, which generally depends on $k_F$. An illustration of the 1D structure is shown in \cref{fig:1d_disp}.  



We now briefly review the procedure of bosonization (see also Ref.~\cite{von1998bosonization}). 
To perform bosonization, we expand the electron operators near the Fermi points,
\ba 
\label{eq:C_p_near_kF}
d_{j,\alpha,p} \approx \gamma_{j,(-1)^{\alpha+1}k_F +p }
\ea 
and introduce the corresponding ``left'' $(\alpha=2)$ and ``right''-moving $(\alpha=1)$ electrons in the conventional bosonization language (see  \cref{fig:1d_disp}) 
\ba 
\label{eq:fourier_fermi}
&
\psi_{j,\alpha }(r)  \approx \sqrt{\frac{2\pi}{L}} \sum_p e^{i p r }d_{j,\alpha,p} ,\quad d_{j,\alpha,p} = \frac{1}{\sqrt{2\pi L}}
\int_{-L/2}^{L/2} 
\psi_{j,\alpha}(r) e^{-ip r} dr 
\ea 
$L =N_y a$ is the length of the wire along the $y$ direction, with $N_y$ the number of sites in each wire and $a$ the lattice spacing. 
We also note that the Fourier transformation convention adopted in \cref{eq:fourier_fermi} differs from that used in Ref.~\cite{von1998bosonization}.
Here, we have taken the continuum limit and used the identity
\ba 
\sum_p e^{ipr} = L \delta(r) ,\quad |r|<L/2
\ea 
The commutation relations are 
\ba 
\{\psi_{j,\alpha}(r) ,\psi_{j',\alpha'}^\dag(r')\} = 2\pi \delta_{j\alpha,j'\alpha'}\delta(r-r')
\ea  

We now briefly review the bosonization procedure. We first introduce $b$ bosons that describe particle-hole fluctuations~\cite{von1998bosonization}
\ba 
b_{j,\alpha,q}^\dag = \frac{i}{\sqrt{n_{q}}}
\sum_k d_{j,\alpha, k+ (-1)^{\alpha+1}q }^\dag d_{j,\alpha,k} ,\quad b_{j,\alpha,q} = \frac{-i}{\sqrt{n_{q}}}
\sum_k d_{j,\alpha, k- (-1)^{\alpha+1}q }^\dag d_{j,\alpha,k}
\ea 
where $q = \frac{2\pi}{L}{n_q}$, with $n_q \in \mathbb{Z}^+$.
We also introduce the filling operator 
\ba 
N_{j,\alpha} = 
\sum_{k}:d_{j,\alpha,k}^\dag d_{j,\alpha,k}:
\ea 
where the $::$ denotes normal ordering with respect to the non-interacting ground state
\ba 
:d_{j,\alpha,k}^\dag d_{j,\alpha,k}:
= d_{j,\alpha,k}^\dag d_{j,\alpha,k} - \Theta\bigg(-(-1)^{\alpha+1}k\bigg) 
\ea 
where $\Theta(x)$ denotes the Heaviside function. 
Combining the $b$ fields and $N$ fields gives the boson field
\ba 
\Phi_{j,\alpha} (r) = - \sum_{ q>0 } \frac{1}{\sqrt{n_q}} 
\bigg( e^{i (-1)^{\alpha+1}q r }b_{j,\alpha,q} 
+ e^{-i
(-1)^{\alpha+1}q r }b^\dag _{j,\alpha,q}\bigg) 
e^{-aq/2} - \frac{2\pi(-1)^{\alpha+1}}{L}N_{j,\alpha}r 
\ea 
The boson fields satisfy the commutation relation
\ba 
\label{eq:phi_phi_commutation}
[\Phi_{j,\alpha}(r),\Phi_{j',\alpha'}(r')] = \delta_{j\alpha,j'\alpha'}(i\pi)(-1)^{\alpha+1}\text{sgn}(r-r') +\mathcal{O}(|r-r'|/L)
\ea 
To recover fermionic statistics in bosonization, we introduce the Klein factor, which satisfies 
\ba 
&\{ F_{j,\alpha}^\dag, F_{j',\alpha'} \} = 2 \delta_{j\alpha,j'\alpha'},\quad 
F_{j,\alpha}^\dag F_{j,\alpha} = F_{j,\alpha} F_{j,\alpha}^\dag =1 ,\quad j\alpha \ne j'\alpha' \nonumber\\
&\{F_{j,\alpha}, F_{j',\alpha'}\} = 
\{F^\dag_{j,\alpha}, F^\dag_{j',\alpha'}\}  = 0\nonumber\\
&[F_{j,\alpha},N_{j',\alpha'}] = -\delta_{j\alpha,j'\alpha'}F_{j,\alpha} 
\ea

We finally reach the following bosonization identity
\ba 
\label{eq:bosonization_identity}
&\psi_{j,\alpha}(r)=F_{j,\alpha}\frac{1}{\sqrt{a}}
e^{-i\Phi_{j,\alpha}(r) }
\nonumber\\ 
&\frac{1}{2\pi}:\psi_{j,\alpha}^\dag(r) \psi_{j,\alpha}(r) : =  (-1)^{\alpha+1}(-1) \frac{1}{2\pi}
\partial_r \Phi_{j,\alpha}(r) 
\ea 
The non-interacting Hamiltonian of each wire, \cref{eq:H_j0}, now reads 
\ba 
\label{eq:H_j0_boson}
H_{j,0} = \int_{-L/2}^{L/2}\frac{dr}{2\pi} 
\sum_\alpha \frac{v}{2} [\partial_r \Phi_{j,\alpha}(r)]^2 
\ea

\section{Sliding Luttinger liquid}
\label{app:SLL}

Within the wire construction, we start from a sliding Luttinger liquid phase. 
The sliding Luttinger liquid phase is described by a free boson theory whose action includes the non-interacting term and forward scattering. 
From \cref{eq:H_j0_boson}, the free Hamiltonian takes the form
\ba 
\label{eq:H_0}
H_0 = \sum_j \int_{-L/2}^{L/2}\frac{dr}{2\pi} 
\sum_\alpha \frac{v}{2} [\partial_r \Phi_{j,\alpha}(r)]^2 
\ea 
The generic forward scattering term can be written as 
\ba 
\label{eq:H_forward}
H_{forward} = 
\int_{-L/2}^{L/2}\frac{dr}{2\pi} 
V_{j-j',\alpha,\alpha'}\partial_r \Phi_{j,\alpha}(r) \partial_r \Phi_{j',\alpha'}(r) 
\ea 
The Hamiltonian of the sliding Luttinger liquid then reads
\ba 
\label{eq:H_SLL_from_H0_Hforward}
H_{SLL} = H_0 + H_{forward}
\ea 
It is also convenient to introduce a new set of fields,
\ba 
\theta_{j}(r) = \frac{1}{{2}}[\Phi_{j,1}(r) + \Phi_{j,2}(r)],\quad 
\phi_{j}(r) = \frac{1}{{2}}[\Phi_{j,1}(r) - \Phi_{j,2}(r)]
\ea 
which obey the commutation relation, from \cref{eq:phi_phi_commutation},
\ba 
[\theta_{j}(r), \phi_{j}(r') ]= \frac{1}{2} i\pi \text{sgn}(r-r') 
\ea 
In the new basis, we find 
\ba 
H_{SLL} 
=&\int_{-L/2}^{L/2}\frac{dr}{2\pi}
\sum_{j,j'} 
\bigg[  v\delta_{j,j'}
\bigg( 
\partial_r \theta_j(r) 
\partial_r \theta_{j'}(r) 
+
\partial_r \phi_j(r) 
\partial_r \phi_{j'}(r) \bigg) 
+ 
 \begin{bmatrix}
\partial_r \theta_j(r) & \partial_r \phi_j(r)     
\end{bmatrix}
\tilde{V}_{j-j'}
 \begin{bmatrix}
\partial_r \theta_{j'}(r) \\ \partial_r \phi_{j'}(r)     
\end{bmatrix}
\bigg] 
\ea 
where 
\ba 
\tilde{V}_{j-j'} = 
\begin{bmatrix}
    1 & 1 \\
    1 & -1 
\end{bmatrix}\cdot V_{j-j'}\cdot \begin{bmatrix}
    1 & 1 \\
    1 & -1 
\end{bmatrix}
\ea 

Finally, the action form of the SLL phase is
\ba 
S_{SLL} = &  2i   
\int_{-L/2}^{L/2} \frac{dr}{2\pi} d\tau 
\sum_{j} \partial_r \phi_{j}(r) 
\partial_\tau \theta_{j}(r)  \nonumber\\
& + \int_\tau 
\int_{-L/2}^{L/2}\frac{dr}{2\pi}
\sum_{j,j'} 
\bigg[  v\delta_{j,j'}
\bigg( 
\partial_r \theta_j(r,\tau) 
\partial_r \theta_{j'}(r,\tau) 
+
\partial_r \phi_j(r)  
\partial_r \phi_{j'}(r) \bigg) 
\nonumber\\
& + \int_\tau \int_{-L/2}^{L/2}\frac{dr}{2\pi }
 \begin{bmatrix}
\partial_r \theta_j(r) & \partial_r \phi_j(r)     
\end{bmatrix}
\tilde{V}_{j-j'}
 \begin{bmatrix}
\partial_r \theta_{j'}(r) \\ \partial_r \phi_{j'}(r)     
\end{bmatrix}
\bigg] 
\ea

\subsection{Interactions}
We now discuss interaction terms. 
Generic interactions beyond forward scattering can be written as~\cite{PhysRevLett.88.036401}
\ba 
\label{eq:fermi_int}
\prod_m [ \psi_{j+m,1}] ^{s_{m}^R}
[\psi_{j+m,2}] ^{s_{m}^L}
\ea 
where $s_{m}^{R/L}$ are integer numbers. Additionally, we let 
\ba 
[\psi_{j,\alpha}]^{s} = \begin{cases}
    \prod_{n=1}^{|s|}\psi_{j,\alpha}(r+\delta r_n) & s> 0 \\ 
    \prod_{n=1}^{|s|}\psi^\dag_{j,\alpha}(r+\delta r_n) & s < 0 
\end{cases}
\ea 
where a negative power indicates a creation operator. A small displacement $\delta r_n$ is introduced to avoid acting with two creation or annihilation operators at the same position. 
The charge conservation of the system imposes a constraint on the allowed interactions 
\ba 
\sum_{m} s_m^R + s_m^L =0 
\ea 
while momentum conservation requires 
\ba 
\sum_m \bigg( s_m^R - s_m^L\bigg) k_F \in  2\pi \mathbb{Z}
\label{eq:momentum_conservation}
\ea 
Because of the lattice structure, momentum conservation is required only modulo $2\pi$.

In terms of the bosonized fields, \cref{eq:fermi_int} can be represented by the generic interaction vertex 
\ba 
\label{eq:bose_int}
V^{\{s_m^R,s_m^L\}}_j(r)  
= e^{-i \sum_m (s_m^R+s_m^L) \theta_{j+m} (r)  
-i \sum_m (s_m^R -s_m^L) \phi_{j+m} (r) 
}
\ea

\section{FCI within the wire construction}
\label{app:FCI_phase_property}

In this section, we briefly review the properties of the FCI phase within the wire construction at filling fraction $\nu=1/3$~\cite{PhysRevLett.88.036401,Meng2020}.

At filling $\nu=1/3$, an FCI instability can develop and gap out the bulk excitations of the SLL phase. 
The FCI phase is stabilized by the following correlated hopping interaction (see Ref.~\cite{PhysRevLett.88.036401} for detailed discussions)
\ba 
\label{eq:FCI_term}
H_{FCI} = 
& \int_{-L/2}^{L/2} \frac{dr}{2\pi} g_{FCI} a^3  
\bigg[ 
 \psi_{j,1}^\dag(r)
[ O_j(r+a)] 
[O_{j+1}(r+a)]  \psi_{j+1,2}(r)
e^{-2ik_F(\delta r_1+\delta r_2)}
+ \text{h.c.}  
\bigg]  \nonumber\\
\approx & \int_{-L/2}^{L/2}\frac{dr}{2\pi  } g_{FCI} \bigg[ [F_{j,1}^\dag]^2 F_{j,2}
F_{j+1,1}^\dag [F_{j+1,2}]^2 
e^{i\bigg[\theta_{j}(r) + 3\phi_{j}(r) - \theta_{j+1}(r) + 3\phi_{j+1}(r)\bigg] } 
+\text{h.c.}\bigg] 
\ea 
where the particle-hole operator reads 
 \ba 
 O_j(r) = \psi_{j,1}^\dag(r) \psi_{j,2} (r)
 \ea 
and $g_{FCI}$ is the coupling strength. 
Here $a$ denotes a small displacement, which is omitted in the bosonized theory. 
Since $\psi_{j,\alpha}$ carries momentum $(-1)^{\alpha}k_F$, the net oscillating factor of the above coupling is unity. 
\ba 
 e^{-i6k_Fa} = e^{i 2\pi } =1 \, .
\ea 
As a result, this term survives in the long-wavelength limit.

\subsection{Properties of the FCI phase}
We briefly discuss the properties of the FCI phase. 
It is useful to introduce 
\ba 
&\tilde{\Phi}_{j,1}(r) =  2 \Phi_{j,1}(r) - \Phi_{j,2}(r) = \theta_j(r) +3\phi_j(r),\quad \tilde{\Phi}_{j,2}(r) = 2 \Phi_{j,2}(r)-\Phi_{j,1}(r) = \theta_{j}(r) -3\phi_{j}(r)  
\ea 
and 
\ba 
\label{eq:FCI_field}
&{\phi}^{FCI}_j(r) = \frac{1}{2} 
[\tilde{\Phi}_{j,1}(r) -\tilde{\Phi}_{j+1,2}(r)] 
= \frac{1}{2}[\theta_j(r) -\theta_{j+1}(r)  + 3\phi_j(r) +3\phi_{j+1}(r)] \nonumber\\
&{\theta}^{FCI}_j(r) = \frac{1}{6}[\tilde{\Phi}_{j,1}(r) + \tilde{\Phi}_{j+1,2}(r) ]=  
\frac{1}{6} [\theta_j(r) +\theta_{j+1}(r) + 3\phi_j(r) -3\phi_{j+1}(r)]
\ea 
which leads to 
\ba 
&[{\theta}^{FCI}_j(r),{\phi}^{FCI}_{j'}(r')] = \delta_{j,j'}\frac{i\pi}{2}\text{sgn}(r-r') 
\ea 

The inverse transformation reads 
\ba 
&\theta_j(r) = 
\frac{1}{2}
[3{\theta}^{FCI}_j(r) + 3 {\theta}^{FCI}_{j-1}(r) +{\phi}^{FCI}_j(r) - {\phi}^{FCI}_{j-1}(r)]
\nonumber\\
&\phi_j(r) = 
\frac{1}{6}
[3{\theta}^{FCI}_j(r) - 3{\theta}^{FCI}_{j-1}(r) +{\phi}^{FCI}_j(r) + {\phi}^{FCI}_{j-1}(r)]
\ea 

With this new set of fields, the interaction term takes a simple form
\ba 
H_{FCI} = \int_{-L/2}^{L/2}\frac{dr}{2\pi  }
g_{FCI} 
\bigg\{ [F_{j,1}^\dag]^2 F_{j,2}
F_{j+1,1}^\dag [F_{j+1,2}]^2 
e^{i 2 {\phi}^{FCI}_j(r) }
+\text{h.c.}\bigg\} 
\ea 

Taking $g_{{FCI}}<0$, once $H_{{FCI}}$ becomes a relevant perturbation, it drives the system into a gapped ground state characterized by the following pinning pattern of the bosonic fields, 
\ba 
{\phi}^{FCI}_j(r) \in  \pi \mathbb{Z} \, .
\ea  

One property of the FCI phase is the presence of fractionalized excitations. 
We consider the excitation created by $e^{i \theta^{FCI}_j(r)}$. From the commutation relation, we find 
\ba 
e^{-i {\theta}^{FCI}_j(r)} 
\phi^{FCI}_{j'}(r') e^{i {\theta}^{FCI}_{j}(r)}
=\phi^{FCI}_{j'}(r') 
- \delta_{j,j'}  \frac{\pi}{2}\text{sgn}(r-r') 
\ea 
Therefore, we consider a FCI ground state described by 
\ba 
 |{\phi}^{FCI}_0\rangle 
 \ea 
 for which 
 \ba 
{\phi}^{FCI}_j(r) |\phi^{FCI}_0\rangle ={ \phi}^{FCI}_0 |\phi^{FCI}_0\rangle  ,\quad \phi_0 \in \pi \mathbb{Z}
 \ea
Acting with $e^{i\theta^{FCI}}$ on the ground state creates a domain wall in the $\phi^{FCI}$ fields
\ba 
\label{eq:domain_wall_prop}
{\phi}^{FCI}_j(r) 
\bigg( e^{i{\theta}^{FCI}_j(r)} 
|\phi_0^{FCI}\rangle \bigg) 
= 
\bigg[ {\phi}^{FCI}_0 
-\frac{\pi}{2} \text{sgn}(r-r') 
\bigg] \bigg( e^{i \tilde{\theta}_j(r)}|\phi^{FCI}_0\rangle\bigg)  
\ea 
Therefore, using \cref{eq:domain_wall_prop}, the charge carried by such a domain wall is 
\ba 
Q \bigg( e^{i{\theta}^{FCI}_j(r)} 
|\phi_0^{FCI}\rangle \bigg)  =& \int \frac{dr}{2\pi} \sum_{\alpha,j} :\psi_{j,\alpha }^\dag(r) \psi_{j,\alpha }(r)  \bigg( e^{i{\theta}^{FCI}_j(r)} 
|\phi_0^{FCI}\rangle \bigg) \nonumber\\ 
\approx  & - 
\int_r \frac{2 dr}{2\pi} \sum_{ j'} \partial_r \phi_{j'}(r)  \bigg( e^{i{\theta}^{FCI}_j(r)} 
|\phi_0^{FCI}\rangle \bigg) \nonumber\\ 
\approx  & \frac{1}{3} \bigg( e^{i{\theta}^{FCI}_j(r)} 
|\phi_0^{FCI}\rangle \bigg) 
\ea 
which is the expected charge $1/3$.

\subsection{Edge mode}
We also analyze the edge mode. 
We start from the commutation relation of the bosonic fields 
\ba 
[ \Phi_{j_1,\alpha_1}(r),\Phi_{j_2,\alpha_2}(r')] = i\pi 
[K_0]_{j_1\alpha_1,j_2\alpha_2} 
\text{sgn}(r-r')
\ea 
where the matrix $K_0$ encodes the commutation relation 
\ba 
[K_0]_{j_1\alpha_1,j_2\alpha_2} = \delta_{j_1\alpha_1,j_2\alpha_2}(-1)^{\alpha_1+1}
\ea 
The FCI field is characterized by the vector $l_j$ 
\ba 
2\phi_j^{FCI}(r) = \sum_{j',\alpha'} [l_j]_{j',\alpha'}\Phi_{j',\alpha'}(r)  
\ea 
where 
\ba 
[l_j]_{j',\alpha'} = 
2\delta_{j',j}\delta_{\alpha',1}
-\delta_{j',j}\delta_{\alpha',2}
-2\delta_{j',j+1}\delta_{\alpha',2} 
+\delta_{j',j+1}\delta_{\alpha',1}
\ea 
It is straightforward to show that 
\ba 
l_j^T \cdot K_0 \cdot l_{j'} = 0 
\ea 
indicating the commuting nature of the FCI fields. 

We next impose open boundary conditions for $N_x$ wires and examine the edge mode. The condensing fields are then characterized by 
\ba 
L_{OBC} = \{l_1,...,l_{N_x-1}\} 
\ea 
where the term crossing the boundary must be cut. 
There are additional boson modes that commute with the condensing modes. We thus seek a mode characterized by $\eta$, with $\Phi^{edge}(r) = \sum_{l',\alpha'}[\eta]_{l'\alpha'}\Phi_{l'\alpha'}(r)$, satisfying 
\ba 
\eta^T \cdot K_0 \cdot l_j 
=0,\quad \forall l_j \in L_{OBC} 
\ea 
To obtain a well-defined vertex operator describing the edge, $\eta$ must be an integer vector.

For the FCI case, such modes can be obtained directly
\ba 
\Phi^{edge}_1(r) = &\Phi_{j=1,1}(r) - 2\Phi_{j=1,2}(r) \nonumber\\
\Phi^{edge}_2(r) = & 2\Phi_{j={N_x-1},1}(r)  - \Phi_{j={N_x-1},2}(r) 
\ea 
Their commutation relations are 
\ba 
[\Phi_j^{edge}(r) ,\Phi_{j'}^{edge}(r')] 
=[K^{edge}]_{jj'}i\pi \text{sgn}(r-r') 
\ea 
where 
\ba 
[K^{edge}] = 
\begin{bmatrix}
    -3 & \\
    & 3 
\end{bmatrix}
\ea 
It is usually convenient to introduce the normalized fields 
\ba 
\theta_1^{edge}(r) = \frac{1}{3}\Phi_1^{edge},\quad \theta_2^{edge}(r) = \frac{1}{3}\Phi_2^{edge}
\ea 

We now examine the edge Hamiltonian. The commutation relation uniquely fixes the Berry-phase part to be 
\ba 
S_0^{edge} = \int \frac{dr}{2\pi} \frac{1}{2} \sum_{j=1,2} (-1)^{j}3\partial_\tau \theta_j^{edge}(r,\tau) \partial_r \theta_j^{edge}(r,\tau)  
\ea 
There is also a non-interacting part, which can be written generically as 
\ba 
S_{free}^{edge} 
= \int \frac{dr}{2\pi}\sum_j \frac{v_j}{2} 
[\partial_r\theta_j^{edge}(r,\tau)]^2 
\ea 
Together, these terms give the conventional edge theory $S_0^{edge}+S_{free}^{edge}$ of the chiral boson.

\subsection{aFCI instabilities}
As pointed out in Ref.~\cite{ShavitOreg2024}, another term closely related to the FCI term can exist in lattice Chern bands. This term is absent in the conventional continuum electron-gas system (see also \cref{eq:momentum_conservation}). The corresponding interaction that stabilizes the aFCI phase reads 
\ba 
\label{eq:aFCI_term}
H_{aFCI} 
\approx & \int_{-L/2}^{L/2}\frac{dr}{2\pi  } g_{aFCI} \bigg[ [F_{j,1}^\dag] [F_{j,2}]^2
[F_{j+1,1}^\dag]^2 [F_{j+1,2}] 
e^{i\bigg[-\theta_{j}(r) + 3\phi_{j}(r) + \theta_{j+1}(r) + 3\phi_{j+1}(r)\bigg] } 
+\text{h.c.}\bigg] 
\ea

\section{Microscopic origin of the FCI interactions and aFCI interactions}
\label{app:FCI_interaction}

In this section, we discuss the microscopic origin of the FCI and aFCI interactions.

We start from on-site repulsions between electrons in different orbitals. In the original orbital basis, we have 
\ba 
H_{U} = \frac{1}{N}\sum_{k,k',p,j,\alpha}
U c_{j, k+p,\alpha}^\dag c_{j,k,\alpha} c_{j,k',3-\alpha}^\dag c_{j,k'+p,3-\alpha}
\ea 
where $j$ is the wire index, $k,p,k'$ are momenta along the $y$ direction, and $\alpha$ denotes the orbital component. 

We then project the electrons to the band-basis operators near the Fermi surface. 
The band-basis operator is defined as (see also \cref{eq:band_basis_operator})
\ba 
\gamma_{j,k}^\dag = \sum_\alpha u_{k,\alpha}c^\dag_{j,k,\alpha}
\ea 
with $u_{p,\alpha}$ the Bloch wave function. 
We are particularly interested in the following channels, which are relevant for FCI phase formation \cite{Meng2020}
\ba 
H_U^{proj} \approx \frac{1}{N}\sum_{p_1,p_2,p_3,\alpha,\alpha'}\tilde{U}^{p_1,p_2,p_3}_{\alpha,\alpha'} \gamma_{j, (-1)^{\alpha+1}k_F+p_1}^\dag 
\gamma_{j,(-1)^{\alpha+1} k_F+p_2}^\dag 
\gamma_{j, (-1)^{\alpha}k_F+p_3} 
c_{j, 3(-1)^{\alpha+1}k_F+p_4,\alpha'}
\delta_{p_1+p_2,p_3+p_4}+\text{h.c.}
\ea 
where $p_1,p_2,p_3,p_4$ denote small momenta near the Fermi surface. 
We note that the two creation operators are assumed to be near the Fermi point $(-1)^{\alpha+1}k_F$, while one of the annihilation operators is taken near $(-1)^{\alpha}k_F$. Momentum conservation then requires the remaining annihilation operator to carry momentum $3(-1)^{\alpha+1}k_F$. Since this momentum is generally far from the low-energy Fermi points, this operator cannot be represented by the low-energy fermion fields and is retained as the original electron operator $c$. 

The projected interaction takes the form
\ba 
\label{eq:def_effective_U}
\tilde{U}^{p_1,p_2,p_3,p_4}_{\alpha,\alpha'} \approx U\sum_\gamma  u^*_{(-1)^{\alpha+1}k_F+p_1,\gamma} 
u^*_{(-1)^{\alpha+1}k_F+p_2,3-\gamma}
[\delta_{\gamma,\alpha'}
u_{(-1)^{\alpha}k_F+p_3,3-\gamma}
-\delta_{\gamma,3-\alpha'}u_{(-1)^{\alpha}k_F+p_3,\gamma}] 
\ea 

We now transform the $\gamma$ electrons into real-space electron operators. 
Approximately, we have 
\ba 
\gamma_{j,(-1)^{\alpha+1}k_F+p} 
\approx \frac{1}{\sqrt{2\pi L}}
\int dr \psi_{j,\alpha}(r)e^{-ipr} 
\ea 
This leads to 
\ba 
H_U^{proj} 
\approx 
\frac{1}{(2\pi L)^{3/2}N}\sum_{p_1p_2p_3,\alpha\alpha'}\int_{r_1r_2r_3}\tilde{U}_{\alpha\alpha'}^{p_1,p_2,p_3} 
\psi_{j,\alpha}^\dag(r_1)\psi_{j,\alpha}^\dag(r_2)\psi_{j,3-\alpha}(r_3) 
c_{j,3(-1)^{\alpha+1}k_F+p_4,\alpha'}
\delta_{p_1+p_2,p_3+p_4}
e^{ip_1r_1+ip_2r_2-ip_3r_3}
\ea 
We consider a gradient expansion of the real-space interaction vertex. Approximately, we expand 
\ba 
\tilde{U}^{p_1p_2p_3}_{\alpha\alpha'} \approx \tilde{U}^{000}_{\alpha\alpha'} + [p_1\partial_{p_1}\tilde{U}^{000}_{\alpha\alpha'} 
+p_2\partial_{p_2}\tilde{U}^{000}_{\alpha\alpha'}]
\ea 
where the leading-order contribution vanishes because of fermionic anticommutation. The leading nonvanishing contribution thus comes from $p_1\partial_{p_1}\tilde{U}^{000}_{\alpha\alpha'} 
+p_2\partial_{p_2}\tilde{U}^{000}_{\alpha\alpha'}$. In real space, this behaves as 
\ba 
H_U^{proj} 
\approx &
\frac{L^2}{(2\pi L)^{3/2}N}\sum_{p,\alpha\alpha'}\int_{r}[(\partial_{p_1}\tilde{U}^{000}_{\alpha\alpha'}i\partial_{r})
\psi_{j,\alpha}^\dag(r)\psi_{j,\alpha}^\dag(r) 
+
\psi_{j,\alpha}^\dag(r)(\partial_{p_2}\tilde{U}^{000}_{\alpha\alpha'}i\partial_{r})\psi_{j,\alpha}^\dag(r) 
] \psi_{j,3-\alpha}(r) 
 \nonumber\\
&c_{j,3(-1)^{\alpha+1}k_F+p,\alpha'}
e^{ipr} \nonumber\\ 
\approx 
&
\frac{L^2}{(2\pi L)^{3/2}N}\sum_{p,\alpha\alpha'}\int_{r}\bigg[(\partial_{p_1}\tilde{U}^{000}_{\alpha\alpha'}i\partial_{r})
-(\partial_{p_2}\tilde{U}^{000}_{\alpha\alpha'}i\partial_{r})\bigg] 
\psi_{j,\alpha}^\dag(r)\psi_{j,\alpha}^\dag(r) \psi_{j,3-\alpha}(r) 
 c_{j,3(-1)^{\alpha+1}k_F+p,\alpha'}
e^{ipr} 
\ea 
We further approximate the real-space derivative by a small displacement characterized by the UV cutoff $a$, and find 
\ba 
\label{eq:H_U_proj_form_v0}
H_U^{proj}\approx  &
\frac{L^2}{(2\pi L)^{3/2}N}\sum_{p,\alpha\alpha'}\int_{r}\frac{1}{a}\bigg[(\partial_{p_1}\tilde{U}^{000}_{\alpha\alpha'}i)
-(\partial_{p_2}\tilde{U}^{000}_{\alpha\alpha'}i)\bigg] 
\psi_{j,\alpha}^\dag(r+a)\psi_{j,\alpha}^\dag(r) \psi_{j,3-\alpha}(r) 
 c_{j,3(-1)^{\alpha+1}k_F+p,\alpha'}
e^{ipr} 
\ea 
We now explicitly evaluate $(\partial_{p_1}\tilde{U}^{000}_{\alpha\alpha'}i)
-(\partial_{p_2}\tilde{U}^{000}_{\alpha\alpha'}i)$, with gauge choice given in \cref{eq:band_basis_operator}. We first notice that (for $\nu=1/3$)
\ba 
&u_{ \pm k_F,\gamma} = 
\begin{bmatrix}
    - \frac{\sqrt{3}M}{\sqrt{6M^2 + 2t_y(t_y\mp \sqrt{3M^2+t_y^2})}}
    & 
         \frac{\mp t_y+\sqrt{3M^2+t_y^2}}{\sqrt{6M^2 + 2t_y(t_y\mp \sqrt{3M^2+t_y^2})}}
\end{bmatrix}_\gamma 
\ea 
In practice, to ensure that only two Fermi points cross the Fermi level at each $k_x$ for all fillings, we generically require $M\ge \frac{1}{\sqrt{2}}t_y$. 
Without loss of generality, we may take $M/t_y \sim 1$, where we find $|u_{k_F,1}|^2/|u_{k_F,2}|^2 = |u_{-k_F,2}|^2/|u_{-k_F,1}|^2\approx 3$. This indicates that the electron near the Fermi energy is predominantly in one orbital flavor. We thus approximately let
\ba 
\label{eq:orbital_weight_near_FS}
u_{  k_F,\gamma}\approx - \delta_{\gamma,1},\quad u_{  -k_F,\gamma}\approx  \delta_{\gamma,2}
\ea 
As for the derivative part, we need to evaluate 
\ba 
\label{eq:partial_particle_link}
W_{ \gamma}(p) = \partial_{p}u^*_{p,\gamma} u^*_{p,3-\gamma}
-u^*_{p,\gamma} \partial_{p} u^*_{p,3-\gamma} = a (-1)^{\gamma+1}\frac{Mt_y/2}{(M^2-t_y^2)\cos(pa)-(M^2+t_y^2) }
\ea 
and 
\ba 
W_\gamma(\pm k_F) \approx a (-1)^{\gamma+1} \frac{-Mt_y}{3M^2+t_y^2}
\ea 
We then conclude from \cref{eq:def_effective_U,eq:orbital_weight_near_FS,eq:partial_particle_link} that
\ba 
\label{eq:simplified_gradient_int_vert}
(\partial_{p_1}\tilde{U}^{000}_{\alpha\alpha'}i)
-(\partial_{p_2}\tilde{U}^{000}_{\alpha\alpha'}i) = 
iU W_1(k_F)  
2\delta_{\alpha,\alpha'}
\ea 

Written in a compact format (by combining \cref{eq:H_U_proj_form_v0,eq:simplified_gradient_int_vert}), we have 
\ba 
\label{eq:H_U_proj_form}
H_U^{proj} 
\approx 
\frac{U}{(2\pi)^{3/2}\sqrt{a}\sqrt{N}}
2i W_1(k_F) \sum_\alpha 
\sum_p \int_r  \psi_{j,\alpha}^\dag(r+a) \psi_{j,\alpha}^\dag(r) 
\psi_{j,3-\alpha}(r) 
c_{j,3(-1)^{\alpha+1}k_F+p_4,\alpha} e^{ip r }+\text{h.c.}
\ea 


We then combine the projected interaction (\cref{eq:H_U_proj_form}) with the inter-wire hopping
\ba 
H_{inter} = \sum_p [t_{x} 
c_{j,p,1}^\dag c_{j+1,p,2} 
+ t_{x'} c_{j,p,1}^\dag c_{j-1,p,2} + \text{h.c.}]
\ea 
Through a cumulant expansion, the third-order contribution to the action gives 
\ba 
S_{eff} \approx & 
\frac{1}{2} 
\int_{\tau_1,\tau_2,\tau_3}
\langle H_U^{proj}(\tau_1) 
H_{inter}^{proj}(\tau_2) 
H_U^{proj}(\tau_3) 
\rangle_>  \nonumber\\
\approx & \int_{\tau_1,\tau_2,\tau_3}\frac{1}{2} 2 
\frac{4U^2|W_1(k_F)|^2}{(2\pi)^3aN}
\sum_{\alpha,p} \int_{r,r'}
\psi_{j,\alpha}^\dag(r+a,\tau_1)\psi_{j,\alpha}^\dag(r,\tau_1) \psi_{j,3-\alpha}(r,\tau_1) 
\psi_{j',\alpha}^\dag(r',\tau_3) 
\psi_{j',3-\alpha}(r',\tau_3) 
\psi_{j',3-\alpha}(r'+a,\tau_3)  \nonumber\\ 
&
e^{ip(r-r') }
\langle c_{j,Q+p,\alpha}(\tau_1)
c_{j,Q+p,\alpha}^\dag(\tau_2)
\rangle 
\langle 
c_{j',Q+p,3-\alpha}(\tau_2)
c_{j',Q+p,3-\alpha}^\dag(\tau_3)
\rangle 
\bigg( 
t_x \delta_{j',j+1}
\delta_{\alpha,1} 
+ t_x' \delta_{j,j'-1}
\delta_{\alpha,2}
\bigg) +\text{h.c.}
\ea 
where $\langle \rangle_>$ indicates integration over high-energy electrons, i.e., electrons near $Q = \pm 3k_F \in 2\pi \mathbb{Z}/a$.  
Let $E_H$ be the energy of electrons near $Q$, away from the Fermi energy. We then find 
\ba 
S_{eff} 
\approx &
\int_\tau 
\frac{4U^2|W_1(k_F)|^2t_x}{(2\pi)^2 E_H^2}
\int _\tau\frac{dr}{2\pi} 
\psi_{j,1}^\dag(r+a,\tau)\psi^\dag_{j,1}(r,\tau) \psi_{j,2}(r,\tau) 
\psi_{j+1,1}^\dag(r,\tau) 
\psi_{j+1,2}(r,\tau)
\psi_{j+1,2}(r+a,\tau) 
+\text{h.c.}\nonumber\\ 
&+\int_\tau 
\frac{4U^2|W_1(k_F)|^2t_x'}{(2\pi)^2 E_H^2}
\int _\tau\frac{dr}{2\pi} 
\psi_{j,1}^\dag(r+a,\tau)\psi^\dag_{j,1}(r,\tau) \psi_{j,2}(r,\tau) 
\psi_{j-1,2}^\dag(r,\tau) 
\psi_{j-1,1}(r,\tau)
\psi_{j-1,1}(r+a,\tau)  +\text{h.c.}
\ea 
The factor $W_1(k_F)$ is related to the quantum geometry of the effective one-dimensional system along the $y$ direction. 
From the definition in \cref{eq:partial_particle_link}, we observe the equality
\ba 
|W_1(p)|^2 = Q^{yy}(p)
\ea 
with $Q^{yy}(p)$ the quantum geometry of the wire at momentum $p$. 
Therefore, the effective coupling reads
\ba 
S_{eff} 
\approx &
\int_\tau 
\frac{4U^2Q^{yy}(k_F)t_x}{(2\pi)^2 E_H^2}
\int _\tau\frac{dr}{2\pi} 
\psi_{j,1}^\dag(r+a,\tau)\psi^\dag_{j,1}(r,\tau) \psi_{j,2}(r,\tau) 
\psi_{j+1,1}^\dag(r,\tau) 
\psi_{j+1,2}(r,\tau)
\psi_{j+1,2}(r+a,\tau) 
+\text{h.c.}\nonumber\\ 
&+\int_\tau 
\frac{4U^2Q^{yy}(k_F)t_x'}{(2\pi)^2 E_H^2}
\int _\tau\frac{dr}{2\pi} 
\psi_{j,1}^\dag(r+a,\tau)\psi^\dag_{j,1}(r,\tau) \psi_{j,2}(r,\tau) 
\psi_{j-1,2}^\dag(r,\tau) 
\psi_{j-1,1}(r,\tau)
\psi_{j-1,1}(r+a,\tau)  +\text{h.c.}
\ea

In the limit closest to the LLL, we have $t_x'=0$. This produces the conventional interactions that stabilize the FCI phase, with 
\ba 
H_{FCI} = \int\frac{dr}{2\pi} 
g_{FCI} a^3\psi_{j,1}^\dag(r+a)\psi^\dag_{j,1}(r) \psi_{j,2}(r)
\psi_{j+1,1}^\dag(r)
\psi_{j+1,2}(r)
\psi_{j+1,2}(r+a) 
+\text{h.c.}
\ea 
where
\ba 
 g_{FCI} \propto   \frac{U^2 Q^{yy}(k_F)t_x}{ E_H^2 }
\ea 

In terms of the boson fields, this gives 
\ba 
\int \frac{dr}{2\pi} 
g_{FCI}
\bigg[ [F_{j,1}^\dag]^2 F_{j,2}
F_{j+1,1}^\dag [F_{j+1,2}]^2 
e^{ i [\theta_{j}(r) + 3\phi_j(r)
-\theta_{j+1}(r) +3\phi_{j+1}(r) ] }
+\text{h.c.}\bigg]
\ea 

However, a finite $t_x'$ leads to another coupling, which we call aFCI, taking the form
\ba 
H_{aFCI} 
=\int\frac{dr}{2\pi} 
g_{aFCI} \psi_{j,1}^\dag(r+a)\psi^\dag_{j,1}(r) \psi_{j,2}(r)
\psi_{j+1,1}^\dag(r)
\psi_{j+1,2}(r)
\psi_{j+1,2}(r+a) 
+\text{h.c.}
\ea 
with
\ba 
g_{aFCI} \propto  \frac{U^2 Q(k_F)t_x'}{ E_H^2 }
\ea 

We remark that the existence of $g_{aFCI}$ is directly related to the quantum geometry of the system. When the system develops a strong quantum geometry as $|t_x'|$ approaches $|t_x|$, $g_{aFCI}$ naturally emerges and can become comparable in strength to $g_{FCI}$, making both FCI and aFCI correlations relevant perturbations.

\section{Perturbative renormalization group analysis}
\label{app:RG}

Having established that both FCI and aFCI perturbations can be relevant to the low-energy physics of a system with strong quantum geometry, we study their interplay using perturbative RG.

\subsection{Superconducting and charge-density-wave correlations induced by FCI and aFCI couplings}
We first focus on the effect of FCI and aFCI scattering. 
From \cref{eq:FCI_term,eq:aFCI_term}, the relevant couplings are 
\ba 
& \int_{-L/2}^{L/2}\frac{dr}{2\pi  } \bigg[ g_{FCI}  
\kappa_{FCI,j} e^{i\Theta_j^{FCI}(r)}
+ g_{aFCI}\kappa_{aFCI,j} e^{i \Theta_j^{aFCI}(r)} 
+\text{h.c.}\bigg] 
\ea 
where the corresponding bosonic fields are 
\ba 
&\Theta_j^{FCI}(r) = \theta_{j}(r) + 3\phi_{j}(r) - \theta_{j+1}(r) + 3\phi_{j+1}(r)\nonumber\\
&\Theta_j^{aFCI}(r) =-\theta_{j}(r) + 3\phi_{j}(r) + \theta_{j+1}(r) + 3\phi_{j+1}(r)
\ea 
and the Klein factors are 
\ba 
&\kappa_{FCI,j}=[F_{j,1}^\dag]^2 F_{j,2}
F_{j+1,1}^\dag [F_{j+1,2}]^2 \nonumber\\
&\kappa_{aFCI,j}=[F_{j,1}^\dag] [F_{j,2}]^2
[F_{j+1,1}^\dag]^2 [F_{j+1,2}] 
\ea 
When the system develops both FCI and aFCI correlations, additional correlations are naturally generated in both the superconducting and charge-density-wave channels. 
This can be seen from the operator product expansion~\cite{cardy1996scaling}
\ba 
: e^{i\Theta_j^{FCI}(R-\frac{r}{2})}:
:e^{i\Theta_j^{aFCI}(R+\frac{r}{2})}: 
\sim 
\frac{1}{|r|^{\Delta^{FCI}+\Delta^{aFCI}-\Delta^{CDW}}}e^{i\Theta_j^{CDW}(R)} \nonumber\\ 
: e^{i\Theta_j^{FCI}(R-\frac{r}{2})}:
:e^{-i\Theta_j^{aFCI}(R+\frac{r}{2})}: 
\sim 
\frac{1}{|r|^{\Delta^{FCI}+\Delta^{aFCI}-\Delta^{SC}}}e^{i\Theta_j^{SC}(R)}
\ea 
where we have introduced 
\ba 
&\Theta_j^{CDW}(r) = \Theta_j^{FCI}(r) + \Theta_j^{aFCI}(r) = 6\phi_j(r) + 6\phi_{j+1}(r)  \nonumber\\
&\Theta_j^{SC}(r) = \Theta_j^{FCI}(r) - \Theta_j^{aFCI}(r) = 2\theta_j(r) -2\theta_{j+1}(r) 
\ea 
where $\Delta^\alpha$ is the scaling dimension of $\Theta_j^\alpha(r)$.

To identify the nature of $\Theta^{CDW}$ and $\Theta^{SC}$, we consider the fermionic representation of the corresponding fields
\ba 
&e^{i\Theta^{CDW}_j} \sim O_j^3 O_{j+1}^3 ,\quad O_j = \psi_{j,1}^\dag \psi_{j,2}
\nonumber\\
&e^{i\Theta^{SC}_j} \sim \Delta_j^\dag \Delta_{j+1}
,\quad 
\Delta_j^\dag = \psi_{j,1}^\dag \psi_{j,2}^\dag
\ea 
Thus $e^{i\Theta^{CDW}_j}$ describes phase locking between particle-hole operators on neighboring wires and stabilizes a CDW phase. By contrast, $e^{i\Theta^{SC}_j}$ takes the form of a Josephson coupling and describes phase locking between pairing fields on neighboring wires, thereby stabilizing an SC phase.

We therefore establish that, when both $g_{FCI}$ and $g_{aFCI}$ are present, correlations in the CDW and SC channels naturally emerge.
The corresponding interaction term can be written as 
\ba 
& \int_{-L/2}^{L/2}\frac{dr}{2\pi  } \bigg[ g_{SC}  
\kappa_{SC,j} e^{i\Theta_j^{SC}(r)}
+ g_{CDW}\kappa_{CDW,j} e^{i \Theta_j^{CDW}(r)} 
+\text{h.c.}\bigg] 
\ea 
where the Klein factors read
\ba 
&\kappa^{CDW}_j = [F_{j,1}^\dag F_{j,2}]^3
[F_{j+1,1}^\dag F_{j+1,2}]^3\nonumber\\
&\kappa^{SC}_j = F_{j,1}^\dag F_{j,2}^\dag 
F_{j+1,2} F_{j+1,1}
\ea

The final interactions considered below are 
\ba 
\label{eq:H_int_on_SLL}
H_{int} = 
\int_{-L/2}^{L/2} \frac{dr}{2\pi}
\sum_{\lambda \in \{FCI, aFCI, CDW,SC\} } \sum_j 
g_{\lambda } \kappa_{\lambda,j}
e^{i\Theta_j^{\lambda}(r)} + \text{h.c.} 
\ea 
with 
\ba 
\label{eq:def_boson_fields_all}
&\Theta_j^{FCI}(r) = \theta_{j}(r) + 3\phi_{j}(r) - \theta_{j+1}(r) + 3\phi_{j+1}(r)\nonumber\\
&\Theta_j^{aFCI}(r) =-\theta_{j}(r) + 3\phi_{j}(r) + \theta_{j+1}(r) + 3\phi_{j+1}(r)
\nonumber\\
&\Theta_j^{SC}(r) = \Theta_j^{FCI}(r) -  \Theta_j^{aFCI}(r) = 2\theta_j(r) -2\theta_{j+1}(r) 
\nonumber\\ 
& \Theta_j^{CDW}(r) = \Theta_j^{FCI}(r) + \Theta_j^{aFCI}(r) = 6\phi_j(r) + 6\phi_{j+1}(r) 
\ea 
and 
\ba
\label{eq:def_klein_all}
& 
\kappa_{FCI,j}=[F_{j,1}^\dag]^2 F_{j,2}
F_{j+1,1}^\dag [F_{j+1,2}]^2 \nonumber\\
&\kappa_{aFCI,j}=[F_{j,1}^\dag] [F_{j,2}]^2
[F_{j+1,1}^\dag]^2 [F_{j+1,2}]\nonumber\\ 
&\kappa^{SC}_j = F_{j,1}^\dag F_{j,2}^\dag 
F_{j+1,2} F_{j+1,1}\nonumber\\ 
&\kappa^{CDW}_j = [F_{j,1}^\dag F_{j,2}]^3
[F_{j+1,1}^\dag F_{j+1,2}]^3
\ea

\subsection{Perturbative RG calculations}
After establishing that superconducting and charge-density-wave orders arise naturally as instabilities of interacting Chern-band systems, we now carry out a perturbative RG analysis to examine the interplay among the FCI, aFCI, SC, and CDW channels.

Following Ref.~\cite{ShavitOreg2024}, we adopt a two-wire approximation, which is sufficient to capture the interplay among the different phases. 
Within the two-wire model, the system is described by two wires labeled by $j=1,2$. We also impose open boundary conditions, so that the FCI and aFCI phases are stabilized by two independent scattering processes.
In the two-wire limit, the bosonic fields can be written in even and odd bases
\ba 
&\theta_e(r) = \frac{1}{\sqrt{2}}(\theta_1(r) +\theta_2(r)),\quad \theta_o(r) = \frac{1}{\sqrt{2}}(\theta_1(r) - \theta_2(r)) \nonumber\\
&\phi_e(r) = \frac{1}{\sqrt{2}}(\phi_1(r) +\phi_2(r)),\quad \phi_o(r) = \frac{1}{\sqrt{2}}(\phi_1(r) - \phi_2(r))  
\ea 
The SLL Hamiltonian can be written in the compact form
\ba 
\label{eq:two_wire_SLL}
H_{SLL} 
= &\int \frac{dr}{2\pi} 
\bigg\{ 
u_e \bigg[ 
K_e [\partial_r \theta_e(r)]^2 
+ \frac{1}{K_e} [\partial_r \phi_e(r)]^2 
\bigg] 
+u_o \bigg[ 
K_o [\partial_r \theta_o(r)]^2 
+ \frac{1}{K_o} [\partial_r \phi_o(r)]^2 
\bigg]  \nonumber\\
&
+v_1 \partial_r \theta_o(r) 
\partial_r \phi_e(r) 
+ v_2 \partial_r \theta_e(r) 
\partial_r \phi_o(r) 
\bigg\} 
\ea 
where $u_e,u_o$ are velocities and $K_e,K_o$ are the Luttinger parameters of the two channels. The parameters $v_1,v_2$ are additional symmetry-allowed couplings that can be induced during the RG flow. 
Microscopically, we consider an on-site repulsion $(u_0)$ and an inter-wire repulsion $(u_1)$ between electron densities with opposite chirality. This allows us to tune $K_e,K_o$ and gives $v_1=v_2=0$. The interaction in the forward-scattering channels can be written as 
\ba 
\label{eq:specific_forward_scattering}
\int \frac{dr}{2\pi} 
\frac{1}{2} \bigg[u_0\delta_{j,j'} + u_1\delta_{j',j+1}+ u_1\delta_{j',j-1}\bigg] \bigg[\rho_{j,1}(r) \rho_{j',2}(r) 
+ \rho_{j,2}(r) \rho_{j',1}(r) 
\bigg] 
\ea 
where $\rho_{j,\alpha}(r) = :\psi_{j,\alpha}^\dag(r)\psi_{j,\alpha}(r):= 
(-1)^{\alpha} \partial_r\Phi_{j,\alpha}(r)$ denotes the density operator. Combining \cref{eq:H_0,eq:specific_forward_scattering} in terms of the bosonic fields, we obtain 
\ba 
H_{SLL} = 
\int \frac{dr}{2\pi} \bigg[ (v-u_0-2u_1 ) [\partial_r\theta_e(r)]^2 + (v+u_0+2u_1)[\partial_r\phi_e(r)]^2
+ (v-u_0+2u_1)[\partial_r\theta_o(r)]^2 + (v+u_0-2u_1)[\partial_r\phi_o(r)]^2\bigg] 
\ea 
This leads to 
\ba 
&u_e = \sqrt{ v^2 -(u_0+2u_1)^2 },\quad u_o = \sqrt{ v^2 -(u_0-2u_1)^2 } \nonumber\\
&K_e = \sqrt{ \frac{v-u_0-2u_1}{v+u_0+2u_1} },\quad K_o = \sqrt{  \frac{v-u_0+2u_1}{v+u_0-2u_1}} 
\ea 
In general, for repulsive interactions with $u_0>0,u_1>0$, we expect 
\ba 
K_e < 1 
\ea 
while $K_o$ is determined by the sign of $u_0-2u_1$
\ba 
\begin{cases}
    K_o < 1 & u_0> 2u_1 \\ 
    K_o>1 & u_0 < 2u_1 
\end{cases}
\ea



Within the two-wire model, the boson fields in \cref{eq:def_boson_fields_all} read
\ba 
&\Theta^{FCI}(r) = \sqrt{2}\theta_o(r) +3\sqrt{2}\phi_e(r) \nonumber\\
&\Theta^{aFCI}(r) = -\sqrt{2}\theta_o(r) +3\sqrt{2}\phi_e(r) \nonumber\\
&\Theta^{SC}(r) = 2\sqrt{2}\theta_o(r) \nonumber\\
&\Theta^{CDW}(r) =6\sqrt{2}\phi_e(r)  
\ea 
and the Klein factors in \cref{eq:def_klein_all} read 
\ba 
& \kappa^{FCI} = [F_{j=1,1}^\dag]^2F_{j=1,2} 
F_{j=2,1}^\dag  F_{j=2,2}^2 \nonumber\\
& \kappa^{aFCI} = F_{j=1,1}^\dag F_{j=1,2}^2
 [F_{j=2,1}^\dag]^2 F_{j=2,2}\nonumber\\ 
&\kappa^{SC} = F_{j=1,1}^\dag F_{j=1,2}^\dag 
F_{j=2,2} F_{j=2,1} \nonumber\\
&\kappa^{CDW} = [F_{j=1,1}^\dag]^3 [F_{j=1,2}]^3
[F_{j=2,1}^\dag]^3 [F_{j=2,2}]^3
\ea 
We note that the field operators involve only the $\phi_e,\theta_o$ fields. We introduce the compact vector representation
\ba 
\Theta^{\lambda}(r) = \mathcal{V}^\lambda\cdot 
\begin{bmatrix}
    \phi_e(r) \\
    \theta_o(r) 
\end{bmatrix}
\ea 
with $\mathcal{V}^\lambda$ a length-2 vector
\ba 
\mathcal{V}^{FCI} =& \begin{bmatrix}
     3\sqrt{2} & \sqrt{2} 
\end{bmatrix} \nonumber\\
\mathcal{V}^{aFCI} = &\begin{bmatrix}
   3\sqrt{2}& - \sqrt{2}  
\end{bmatrix} \nonumber\\
\mathcal{V}^{SC} =& \begin{bmatrix}
  0 &  2 \sqrt{2} 
\end{bmatrix} \nonumber\\
\mathcal{V}^{CDW} =& \begin{bmatrix}
      6\sqrt{2} &0
\end{bmatrix} 
\ea

We now provide the detailed calculation of the vertex-operator propagators and the operator-product expansion. 
We start from the free-boson Green's function, which takes the following form perturbatively in $v_1,v_2$
\ba 
G_{ij}(q,z) =&  \langle 
\begin{bmatrix}
    \theta_e(q,z) \\ \phi_e(q,z) \\ \theta_o(q,z) 
    \\ 
    \phi_o(q,z) 
\end{bmatrix}_i 
\begin{bmatrix}
    \theta_e(-q,-z) & \phi_e(-q,-z) &  \theta_o(-q,-z)  & 
    \phi_o(-q,-z) 
\end{bmatrix}_j\rangle \nonumber\\
= & \frac{1}{2\pi} 
\begin{bmatrix}
    \frac{\pi u_e/K_e}{-z^2 +(u_eq)^2}  & \frac{\pi z/q}{-z^2 +(u_eq)^2}  
    \\ 
    \frac{\pi z/q}{-z^2 +(u_eq)^2}  & \frac{\pi u_e K_e}{-z^2 + (u_eq)^2} 
    \\
        & & \frac{\pi u_o/K_o}{-z^2 +(u_oq)^2}  & \frac{\pi z/q}{-z^2 +(u_oq)^2} 
    \\ 
   & &  \frac{\pi z/q}{-z^2 +(u_oq)^2}  & \frac{\pi u_o K_o}{-z^2 + (u_oq)^2}
\end{bmatrix}_{ij}   
+ 
\frac{1 }{4[-z^2 + (u_eq)^2] 
[-z^2 + (u_oq)^2]}\nonumber\\
&
\begin{bmatrix}
    0 & 0 & -qz \bigg( 
    \frac{u_o}{K_o}v_1 + \frac{u_e}{K_e}v_2
    \bigg) & 
   -q^2 \frac{u_eu_oK_o}{K_e}v_2 
   -z^2 v_1  \\
   0 & 0& - q^2 \frac{K_e u_e u_o}{K_o}v_1-  z^2 v_2 & - qz\bigg( K_e u_e v_1 + K_ou_ov_2\bigg) 
   \\
      -qz \bigg( 
    \frac{u_o}{K_o}v_1 + \frac{u_e}{K_e}v_2
    \bigg) &  - q^2 \frac{K_e u_e u_o}{K_o}v_1-  z^2 v_2
    & 0 &0 \\
-q^2 \frac{u_eu_oK_o}{K_e}v_2 
   -z^2 v_1& - qz\bigg( K_e u_e v_1 + K_ou_ov_2\bigg) & 0 &0
\end{bmatrix}
\nonumber\\
&+ \mathcal{O}(v_1^2, v_2^2, v_1v_2)
\ea 
where $\theta_\alpha(q,i\omega) =\sqrt{ \frac{2\pi}{L\beta }} \int_{r,\tau} e^{iqr-i\omega \tau}\theta_\alpha(r,\tau) ,\phi_\alpha(q,i\omega) =\sqrt{ \frac{2\pi}{L\beta }} \int_{r,\tau} e^{iqr-i\omega \tau}\phi_\alpha(r,\tau) $. 

The relevant propagators in real space read 
\ba 
\label{eq:boson_prop_real_space_1}
G_{22}(r,\tau) 
=\langle \phi_{e}(r,\tau) \phi_e(0,0)\rangle \approx   & \frac{2\pi }{N_ya\beta}
\sum_{i\Omega,q}G_{22}(q,i\Omega)e^{-i\Omega \tau  +iq r} \nonumber\\
=&\frac{1}{N_ya}\sum_q  \frac{\pi K_e e^{- |u_e q | |\tau| +iq r} }{2 |q| } \approx - \frac{K_e }{2 }
\log\bigg(\frac{2\pi}{L_y}  \sqrt{ |u_e\tau|^2 + r^2 } 
\bigg) \nonumber\\ 
G_{33}(r,\tau) 
=\langle \theta_{o}(r,\tau) \theta_o(0,0)\rangle  \approx & \frac{2\pi}{N_ya\beta}
\sum_{i\Omega,q}G_{33}(q,i\Omega)e^{-i\Omega \tau  +iq r}  \approx  - \frac{1}{2K_o }
\log\bigg( \frac{2\pi}{L_y} \sqrt{ |u_o\tau|^2 + r^2 } 
\bigg) 
\ea 
For the off-diagonal term, we first note that 
\ba 
G_{23}(q,z) = & 
\frac{\pi}{2 K_o}\bigg\{
\bigg[ 
\frac{1}{-z^2+(u_eq)^2}
-\frac{1}{-z^2+(u_oq)^2}]
\bigg] 
\frac{1}{u_o^2 -u_e^2}
\bigg(
-K_eu_eu_ov_1 
\bigg) \nonumber\\
&
+ 
\bigg[ 
\frac{1/u_o^2}{-z^2+(u_eq)^2}
-\frac{1/u_e^2}{-z^2+(u_oq)^2}
\bigg] 
\frac{-1}{\frac{1}{u_o^2}-\frac{1}{u_e^2}} 
\bigg(-K_ov_2\bigg) 
\bigg\} 
\ea 
In real space and imaginary time, this gives 
\ba 
\label{eq:boson_prop_real_space_2}
&G_{23}(r,\tau)= \langle \phi_e(r,\tau) \theta_o(0,0)\rangle  \nonumber\\
\approx 
& -\frac{1}{4K_o} 
\frac{K_eu_ov_1 + K_ou_ev_2}{u_e^2-u_o^2}
\log\bigg( \frac{2\pi}{L_y} \sqrt{ |u_e\tau|^2 + r^2 }
\bigg) 
-\frac{1}{4K_o} 
\frac{K_eu_ev_1 + K_ou_ov_2}{-u_e^2+u_o^2}
\log\bigg( \frac{2\pi}{L_y} \sqrt{ |u_o\tau|^2 + r^2 }
\bigg) 
\ea

We are interested in the normal-ordered vertex
\ba 
: e^{i\Theta^{\lambda}(r,\tau)}: 
= e^{i\Theta^{\lambda}(r,\tau)}
e^{ \frac{1}{2} 
\langle [\Theta^\lambda(r,\tau)]^2\rangle }
\ea 
where the self-contraction term has been removed. 
Using \cref{eq:boson_prop_real_space_1,eq:boson_prop_real_space_2}, the vertex propagators take the following scaling forms~\cite{von1998bosonization}
\ba 
&\langle :e^{i\Theta^{\lambda}(r,\tau)}:
:e^{-i\Theta^\lambda(r',\tau')}:
\rangle
\approx 
e^{ 
\langle \Theta^\lambda(r,\tau) \Theta^{\lambda}(r',\tau')\rangle 
}
\approx 
\bigg[ \frac{1}{\frac{2\pi}{L_y}\sqrt{[u_e|\tau-\tau'|]^2 + |r-r'|^2 }}
\bigg] ^{2\Delta_e^\lambda}
\bigg[ 
\frac{1}{\frac{2\pi}{L_y}\sqrt{[u_o|\tau-\tau'|]^2 + |r-r'|^2 }}
\bigg] ^{2\Delta_o^\lambda} 
\ea 
The scaling dimensions are 
\ba 
\label{eq:general_scaling_dimension}
&2\Delta_e^{\lambda} = [\mathcal{V}^\lambda_1]^2 \frac{K_e}{2}  
+ \mathcal{V}^\lambda_1\mathcal{V}^\lambda_2 \frac{K_eu_ov_1 +K_ou_ev_2}{2K_o(u_e^2-u_o^2)} \nonumber\\ 
&2\Delta_o^{\lambda} = [\mathcal{V}^\lambda_2]^2 \frac{1}{2K_o} 
+ \mathcal{V}^\lambda_1\mathcal{V}^\lambda_2 \frac{K_eu_ev_1 +K_ou_ov_2}{2K_o(u_o^2-u_e^2)} \nonumber\\
&\Delta^\lambda = \Delta_e^\lambda + \Delta_o^\lambda 
\ea 
where we keep terms only to linear order in $v_1,v_2$. 

We now perform a perturbative RG analysis of instabilities of the SLL phase. We consider perturbations of the form given in \cref{eq:H_int_on_SLL}
\ba 
S_{int} = \sum_{\lambda \in \{FCI,aFCI,SC,CDW\} } g_\lambda 
\int_\tau \int \frac{dr}{2\pi} \kappa_\lambda :
e^{i\Theta^{\lambda}(r,\tau)}: +\text{h.c.}
\ea 
with coupling constant $g_\lambda$ and corresponding Klein factor $\kappa_\lambda$. We further introduce the dimensionless coupling $y_\lambda$ \cite{cardy1996scaling}
\ba 
g_\lambda = y_\lambda a^{-2 + \Delta^\lambda } 
\bigg(\frac{2\pi }{L_y}\bigg)^{\Delta^\lambda}
\sqrt{u_eu_o}
\ea 
where $a$ is the UV cutoff.

At leading order, rescaling $a$ via $a \rightarrow ae^{l}$ rescales the coupling constant,
\ba 
\label{eq:tree_level_rescaling}
y_\lambda \rightarrow y_\lambda + y_\lambda \bigg[ 2-\Delta^\lambda 
\bigg] l 
\ea 
which gives the leading-order RG equation and indicates that the perturbation channel $\lambda$ becomes relevant when $2-\Delta^\lambda >0$. 

We next consider the second-order effect, which captures the interplay among FCI, aFCI, CDW, and SC channels. 
At second order, the cumulant expansion gives 
\ba 
\label{eq:cumulant_exp}
S_{correction}^{(2)} = &-\frac{1}{2}\sum_{\lambda,\lambda'}\bigg[ \langle S_\lambda S_{\lambda'} \rangle_>
-\langle S_\lambda \rangle _>\langle S_{\lambda'}\rangle_>\bigg] 
\ea 
where $\langle \rangle_>$ denotes integrating out short-distance fluctuations between the cutoffs $a$ and $a e^l$. 
To evaluate \cref{eq:cumulant_exp}, we ustilize
\ba 
:e^{is\Theta^{\lambda}(R+\frac{r}{2},T+\frac{\tau}{2})}:
:e^{is'\Theta^{\lambda'}(R-\frac{r}{2},T-\frac{\tau}{2})}: 
= e^{- ss' 
\langle 
\Theta^{\lambda}(R+\frac{r}{2},T+\frac{\tau}{2})
\Theta^{\lambda'}(R-\frac{r}{2},T-\frac{\tau}{2})
\rangle }: 
e^{i 
\bigg[ s 
\Theta^{\lambda}(R+\frac{r}{2},T+\frac{\tau}{2})+ s'\Theta^{\lambda'}(R-\frac{r}{2},T-\frac{\tau}{2})
\bigg]
}:
\ea 
with $s,s' \in \{+1,-1\}$. 
We further perform a gradient expansion and keep the leading-order contributions. For $s\Theta^\lambda \ne -s'\Theta^{\lambda'}$, at leading order, we have 
\ba 
\label{eq:normal_order_two_vertex}
:e^{is\Theta^{\lambda}(R+\frac{r}{2},T+\frac{\tau}{2})}:
:e^{is'\Theta^{\lambda'}(R-\frac{r}{2},T-\frac{\tau}{2})}: 
\approx  e^{- ss'
\langle 
\Theta^{\lambda}(R+\frac{r}{2},T+\frac{\tau}{2})
\Theta^{\lambda'}(R-\frac{r}{2},T-\frac{\tau}{2})
\rangle }: 
e^{i 
\bigg[ s
\Theta^{\lambda}(R,T)+ s'\Theta^{\lambda'}(R,T)
\bigg]
}:
\ea 
Using \cref{eq:boson_prop_real_space_1,eq:boson_prop_real_space_2}, the coefficient reads
\ba 
\label{eq:OPE_1}
& e^{ -ss' \langle \Theta^\lambda(R+ \frac{r}{2},T+\frac{\tau}{2}) 
\Theta^{\lambda'}(R-\frac{r}{2},T- \frac{\tau}{2}\rangle  ) 
} \nonumber\\
\approx& 
\bigg[ 
\frac{1}{ \frac{2\pi}{L_y}
\sqrt{ |u_e\tau|^2 + |r|^2 }}
\bigg]^{ -ss'\mathcal{V}^\lambda_1\mathcal{V}_1^{\lambda'} \frac{K_e}{2} 
- ss'(\mathcal{V}^\lambda_1\mathcal{V}_2^{\lambda'} +
\mathcal{V}^\lambda_2\mathcal{V}_1^{\lambda'})\frac{K_eu_ov_1 +K_ou_ev_2}{4K_o(u_e^2-u_o^2)} 
} \nonumber\\
&
\bigg[ 
\frac{1}{ \frac{2\pi}{L_y}
\sqrt{ |u_o\tau|^2 + |r|^2 }}
\bigg]^{ -ss'\mathcal{V}^\lambda_2\mathcal{V}_2^{\lambda'} \frac{1}{2K_o} 
-ss'(\mathcal{V}^\lambda_1\mathcal{V}_2^{\lambda'} +
\mathcal{V}^\lambda_2\mathcal{V}_1^{\lambda'})\frac{K_eu_ev_1 +K_ou_ov_2}{4K_o(u_o^2-u_e^2)} 
} \nonumber\\
\approx & 
\bigg[ 
\frac{1}{ \frac{2\pi}{L_y}
\sqrt{ |u_e\tau|^2 + |r|^2 }}
\bigg]^{ -\Delta_e^{s\lambda+s'\lambda'}+\Delta_e^{s\lambda} +  \Delta_e^{s'\lambda'}
}
\bigg[ 
\frac{1}{ \frac{2\pi}{L_y}
\sqrt{ |u_o\tau|^2 + |r|^2 }}
\bigg]^{-\Delta_o^{s\lambda+s'\lambda'}+ \Delta_o^{s\lambda} +  \Delta_o^{s'\lambda'}}
\ea 
where $\Delta_{e/o}^{s\lambda+s'\lambda'}$ denotes the scaling dimension of $:e^{is\Theta^\lambda(R,T)+s'\Theta^{\lambda'}(R,T) }:$, and $\Delta_{e/o}^{s\lambda}$ denotes the scaling dimension of $:e^{is\Theta^\lambda(R,T)}:$ taking the form of 
\ba 
&2\Delta_{e}^{s\lambda+s'\lambda'} = [sV^\lambda_1  + s' V^{\lambda'}_1]^2 
\frac{K_e}{2}
 + (sV^\lambda_1 + s'V^{\lambda'}_1)
 (sV^\lambda_2 + v'V^{\lambda'}_2)\frac{K_eu_ov_1+K_ou_ev_2}{2K_o(u_e^2-u_o^2)}\nonumber\\ 
 &2\Delta_{o}^{s\lambda+s'\lambda'} = [sV^\lambda_2  + s' V^{\lambda'}_2]^2 
\frac{1}{2K_o}
 + (sV^\lambda_1 + s'V^{\lambda'}_1)
 (sV^\lambda_2 + v'V^{\lambda'}_2)\frac{K_eu_ev_1+K_ou_ov_2}{2K_o(-u_e^2+u_o^2)} \nonumber\\
 & \Delta_{e/o}^{s\lambda} = \Delta^\lambda_{e/o}
\ea 

We next integrate over
\ba 
a<
\sqrt{u_eu_o|\tau|^2 +|r|^2 }< ae^{l} 
\ea 
This leads to 
\ba 
\label{eq:OPE_1_integrate}
&\int_{ a<
\sqrt{u_eu_o|\tau|^2 +|r|^2 }< ae^{l} }d\tau \frac{dr}{2\pi} \bigg[ 
\frac{1}{ \frac{2\pi}{L_y}
\sqrt{ |u_e\tau|^2 + |r|^2 }}
\bigg]^{ -\Delta_e^{s\lambda+s'\lambda'}+ \Delta_e^{s\lambda} +  \Delta_e^{s'\lambda'}
}
\bigg[ 
\frac{1}{ \frac{2\pi}{L_y}
\sqrt{ |u_o\tau|^2 + |r|^2 }}
\bigg]^{-\Delta_o^{{s\lambda+s'\lambda'}}+\Delta_o^{s\lambda }+\Delta_o^{s'\lambda'}}
\nonumber\\
\approx &  
\bigg( \frac{2\pi}{L_y}\bigg)^{\Delta^{{s\lambda+s'\lambda'}}-\Delta^{s\lambda }-\Delta^{s'\lambda'}} \frac{1}{\sqrt{u_eu_o}} 
a^{2 +\Delta^{{s\lambda+s'\lambda'}}-\Delta^{s\lambda}-\Delta^{s'\lambda'}}l 
\mathcal{A}^{s\lambda,s'\lambda'}_{u_e,u_o}
\ea 
where 
\ba 
\mathcal{A}^{s\lambda,s'\lambda'}_{u_e,u_o}
=\frac{1}{2\pi} \int dx \bigg[ 
\frac{1}{
\sqrt{ \frac{u_e}{u_o}\cos^2(x) + \sin^2(x) }}
\bigg]^{ -\Delta_e^{{s\lambda+s'\lambda'}}+ \Delta_e^{s\lambda} +  \Delta_e^{s'\lambda'}
}
\bigg[ 
\frac{1}{ 
\sqrt{ \frac{u_o}{u_e}\cos^2(x) + \sin^2(x)}}
\bigg]^{-\Delta_o^{{s\lambda+s'\lambda'}}+\Delta_o^{s\lambda} +\Delta_o^{s'\lambda'}}
\ea 
In the simplified limit $u_e=u_o=u$, 
\ba 
\mathcal{A}^{s\lambda,s'\lambda'}_{u,u} = 1 
\ea

Another useful operator product expansion comes from \cref{eq:normal_order_two_vertex} with $s=s'=1,\lambda=\lambda'$ . After a gradient expansion, this gives 
\ba 
\label{eq:OPE_2}
&:e^{i\Theta^{\lambda}(R+\frac{r}{2},T+\frac{\tau}{2})}:
:e^{-i\Theta^{\lambda}(R-\frac{r}{2},T-\frac{\tau}{2})}: 
\approx  e^{
\langle 
\Theta^{\lambda}(R+\frac{r}{2},T+\frac{\tau}{2})
\Theta^{\lambda}(R-\frac{r}{2},T-\frac{\tau}{2})
\rangle }: 
e^{i 
\bigg[ 
\Theta^{\lambda}(R+\frac{r}{2},T+\frac{\tau}{2})- \Theta^{\lambda}(R-\frac{r}{2},T-\frac{\tau}{2})
\bigg]
}: \nonumber\\
\rightarrow  & 
\bigg[ \frac{1}{\frac{2\pi}{L_y}\sqrt{[u_e|\tau|]^2 + |r|^2 }}
\bigg] ^{2\Delta_e^\lambda}
\bigg[ 
\frac{1}{\frac{2\pi}{L_y}\sqrt{[|u_o\tau|]^2 + |r|^2 }}
\bigg] ^{2\Delta_o^\lambda} 
\bigg[ 
-\frac{r^2}{2 }
[\partial_r \Theta^\lambda(R,T)]^2 
-\frac{\tau^2}{2 }
[\partial_\tau \Theta^\lambda(R,T)]^2 
\bigg] 
\ea 
where we keep only the leading nonzero contributions.

Integrating over short-distance fluctuations gives 
\ba 
\label{eq:OPE_2_integrate}
&\int_{a<
\sqrt{u_eu_o|\tau|^2 +|r|^2 }< ae^{l}} 
d\tau \frac{dr}{2\pi } 
\bigg[ \frac{1}{\frac{2\pi}{L_y}\sqrt{[u_e|\tau|]^2 + |r|^2 }}
\bigg] ^{2\Delta_e^\lambda}
\bigg[ 
\frac{1}{\frac{2\pi}{L_y}\sqrt{[|u_o\tau|]^2 + |r|^2 }}
\bigg] ^{2\Delta_o^\lambda} r^2 
= \bigg( \frac{2\pi}{L}\bigg)^{-2\Delta^\lambda}
a^{4-2\Delta^\lambda}\frac{1}{\sqrt{u_eu_o}} \mathcal{A}^{\lambda;r}_{u_e,u_o} l 
\nonumber\\
&\int_{a<
\sqrt{u_eu_o|\tau|^2 +|r|^2 }< ae^{l}} 
d\tau \frac{dr}{2\pi } 
\bigg[ \frac{1}{\frac{2\pi}{L_y}\sqrt{[u_e|\tau|]^2 + |r|^2 }}
\bigg] ^{2\Delta_e^\lambda}
\bigg[ 
\frac{1}{\frac{2\pi}{L_y}\sqrt{[|u_o\tau|]^2 + |r|^2 }}
\bigg] ^{2\Delta_o^\lambda} \tau^2 
= \bigg( \frac{2\pi}{L}\bigg)^{-2\Delta^\lambda}
a^{4-2\Delta^\lambda} \frac{1}{\sqrt{u_eu_o}^3}\mathcal{A}^{\lambda;\tau}_{u_e,u_o} l 
\ea 
where 
\ba 
&\mathcal{A}^{\lambda;r}_{u_e,u_o}  = \int \frac{dx}{2\pi} 
\bigg[ \frac{1}{
\sqrt{ \frac{u_e}{u_o}\cos^2(x) + \sin^2(x) }}
\bigg]^{ 2 \Delta_e^\lambda 
}
\bigg[ 
\frac{1}{ 
\sqrt{ \frac{u_o}{u_e}\cos^2(x) + \sin^2(x)}}
\bigg]^{2 \Delta_o^\lambda }\sin^2(x) 
\nonumber\\
&\mathcal{A}^{\lambda;\tau}_{u_e,u_o}  = \int \frac{dx}{2\pi} 
\bigg[ \frac{1}{
\sqrt{ \frac{u_e}{u_o}\cos^2(x) + \sin^2(x) }}
\bigg]^{ 2 \Delta_e^\lambda 
}
\bigg[ 
\frac{1}{ 
\sqrt{ \frac{u_o}{u_e}\cos^2(x) + \sin^2(x)}}
\bigg]^{2 \Delta_o^\lambda }\cos^2(x) 
\ea 
Again, in the $u_e=u_o=u $ limit, we have
\ba 
\mathcal{A}^{\lambda;r}_{u,u}=\mathcal{A}^{\lambda;\tau}_{u,u} = \frac{1}{2} 
\ea

We also comment that 
\ba 
\label{eq:fusion_rule}
&\Theta^{FCI}(r,\tau) + \Theta^{aFCI}(r,\tau) = \Theta^{CDW}(r),\quad 
\Theta^{FCI}(r,\tau) - \Theta^{aFCI}(r,\tau) = \Theta^{SC}(r) \nonumber\\
&\kappa^{FCI} \kappa^{aFCI} = \kappa^{CDW}
,\quad \kappa^{FCI}[\kappa^{aFCI}]^\dag = \kappa^{SC}
\ea 

Collecting all contributions using \cref{eq:cumulant_exp,eq:OPE_1,eq:OPE_1_integrate,eq:OPE_2,eq:OPE_2_integrate,eq:fusion_rule}, we obtain the following correction from integrating over the short-distance modes
\ba 
\label{eq:second_order_correction_full}
&S_{correction}^{(2)} \nonumber\\
=&-
y_{FCI}y_{aFCI} 
\int u d\tau  \int \frac{dr}{2\pi}\kappa_{FCI} \kappa_{aFCI}  :
\bigg( \frac{2\pi}{L_y}\bigg)^{\Delta^{CDW}}
a^{-2 + \Delta^{CDW}} 
:
e^{i\Theta^{CDW}(r,\tau)}:  l  
+\text{h.c.}  \nonumber\\
&-
y_{FCI}y_{aFCI} 
\int u d\tau  \int \frac{dr}{2\pi}\kappa_{FCI} \kappa^\dag_{aFCI}  :
\bigg( \frac{2\pi}{L_y}\bigg)^{\Delta^{SC}}
a^{-2 + \Delta^{SC}} 
:
e^{i\Theta^{SC}(r,\tau)}:  l  
+\text{h.c.}  \nonumber\\
&-
y_{FCI}y_{CDW} 
\int u d\tau  \int \frac{dr}{2\pi}\kappa^\dag_{FCI} \kappa_{CDW}  :
\bigg( \frac{2\pi}{L_y}\bigg)^{\Delta^{aFCI}}
a^{-2 + \Delta^{aFCI}} 
:
e^{i\Theta^{aFCI}(r,\tau)}:  l 
+\text{h.c.}  \nonumber\\
&-
y_{aFCI}y_{CDW} 
\int u d\tau  \int \frac{dr}{2\pi}\kappa^\dag_{aFCI} \kappa_{CDW}  :
\bigg( \frac{2\pi}{L_y}\bigg)^{\Delta^{FCI}}
a^{-2 + \Delta^{FCI}} 
:
e^{i\Theta^{FCI}(r,\tau)}: l 
+\text{h.c.}  \nonumber\\
&-
y_{SC}y_{aFCI} 
\int u d\tau  \int \frac{dr}{2\pi}\kappa_{SC} \kappa_{aFCI}  :
\bigg( \frac{2\pi}{L_y}\bigg)^{\Delta^{FCI}}
a^{-2 + \Delta^{FCI}} 
:
e^{i\Theta^{FCI}(r,\tau)}:  l 
+\text{h.c.}  \nonumber\\
&-
y_{SC}y_{FCI} 
\int u d\tau  \int \frac{dr}{2\pi}\kappa^\dag_{SC} \kappa_{FCI}  :
\bigg( \frac{2\pi}{L_y}\bigg)^{\Delta^{aFCI}}
a^{-2 + \Delta^{aFCI}} 
:
e^{i\Theta^{aFCI}(r,\tau)}:  l 
+\text{h.c.}  \nonumber\\
& 
-
\frac{1}{2}
\sum_\lambda y_\lambda^2 
\int u d\tau \int \frac{dr}{2\pi} 
\frac{-1}{4} 
\bigg[ 
[\partial_r \Theta^\lambda(r,\tau)]^2 + 
[\frac{1}{u}\partial_\tau  \Theta^\lambda(r,\tau)]^2 
\bigg] l +\text{h.c.}
\ea
where, to obtain a simple analytical expression, we focus on the $u_e=u_o=u$ limit and keep only the corrections to $g_{FCI/aFCI/SC/CDW}$ and to the free-boson part.

We can now derive the RG equations explicitly. We first focus on the free-boson part. 
The original action with the $\theta_e,\phi_o$ fields integrated out reads 
\ba 
S_{free,\phi_e\theta_o} \approx &
\int_\tau \int \frac{udr}{2\pi} 
\bigg\{ 
\frac{1}{K_e}  
\bigg[ [ \partial_r \phi_e(r,\tau)]^2 
+  [\frac{1}{u}\partial_\tau \phi_e(r,\tau)]^2\bigg]   
+K_o 
\bigg[ [ \partial_r \theta_o(r,\tau)]^2 
+ [\frac{1}{u}\partial_\tau \theta_o(r,\tau)]^2  \bigg] 
\nonumber\\
& +\frac{1}{u}  v_1 
\partial_r \phi_e(r,\tau) 
\partial_r \theta_o(r,\tau) 
-  \frac{1}{u} v_2\frac{K_o}{K_eu^2}
\partial_\tau \phi_e(r,\tau) 
\partial_\tau \theta_o(r,\tau) 
\bigg\}  
\ea 
The correction to the free-boson part reads, from \cref{eq:second_order_correction_full},
\ba 
S_{free,correction} 
\approx &
\frac{ l}{4} \int ud\tau \int \frac{dr}{2\pi} 
 \bigg(72 y_{CDW}^2 
+ 18 y_{FCI}^2 +18 y_{aFCI}^2 \bigg)
\bigg( [\partial_r \phi_e(r,\tau)]^2 +[\frac{1}{u}\partial_\tau \phi_e(r,\tau)]^2 \bigg) 
 \nonumber\\
& + 
\frac{ l}{4} \int ud\tau \int \frac{dr}{2\pi} 
 \bigg( 8 y_{SC}^2 
+ 2 y_{FCI}^2 +2 y_{aFCI}^2 \bigg)
\bigg( [\partial_r \theta_o(r,\tau)]^2 +[\frac{1}{u}\partial_\tau \theta_o(r,\tau)]^2 \bigg) 
\nonumber\\
& + 
\frac{ l}{4} \int ud\tau \int \frac{dr}{2\pi} 
 2\bigg( 6y_{FCI}^2 -6y_{aFCI}^2\bigg) 
\bigg( [\partial_r \theta_o(r,\tau)] 
[\partial_r \phi_e(r,\tau)]
+ [\frac{1}{u}\partial_\tau \theta_o(r,\tau)] 
[\frac{1}{u} \partial_\tau \phi_e(r,\tau)]
\bigg) 
\ea 
We can thus introduce the renormalized parameters $\tilde{v}_1,\tilde{v}_2,\tilde{K}_e,\tilde{K}_o$
\ba 
&\frac{1}{\tilde{K_e}} = \frac{1}{K_e} + \frac{l}{4} \bigg( 72y_{CDW}^2 + 18y_{FCI}^2 + 18y_{aFCI}^2\bigg) \nonumber\\
&\tilde{K_o} = K_o + \frac{l}{4} \bigg( 8 y_{SC}^2 
+ 2 y_{FCI}^2 +2 y_{aFCI}^2 \bigg) \nonumber\\ 
& \tilde{v_1} = v_1+ \frac{l}{2} \bigg(6y_{FCI}^2-6 y_{aFCI}^2\bigg)  u\nonumber\\ 
& \tilde{v_2} =v_2 -\frac{l}{2} 
\bigg(6y_{FCI}^2-6 y_{aFCI}^2\bigg) 
\frac{K_e}{K_o} u
\ea 

This gives the RG equations 
\ba 
\label{eq:RG_flow_K}
&\partial_l\frac{1}{K_e} =  \frac{1}{4} \bigg( 72y_{CDW}^2 + 18y_{FCI}^2 + 18y_{aFCI}^2\bigg) \nonumber\\
&\partial_l K_o= \frac{1}{4} \bigg( 8 y_{SC}^2 
+ 2 y_{FCI}^2 +2 y_{aFCI}^2 \bigg) \nonumber\\ 
& \partial_l v'_1 = \frac{1}{2} \bigg(6y_{FCI}^2-6 y_{aFCI}^2\bigg)  \nonumber\\ 
& \partial_l v'_2= -\frac{1}{2} 
\bigg(6y_{FCI}^2-6 y_{aFCI}^2\bigg) 
\frac{K_e}{K_o} 
\ea 
where $v'_i = v_i/u$. 

In addition, the renormalization of the coupling constants can be inferred from \cref{eq:tree_level_rescaling,eq:second_order_correction_full}
\ba 
\label{eq:RG_flow_y}
&\partial_l y_{FCI} = (2-\Delta^{FCI})y_{FCI}
-  (y_{aFCI}y_{CDW} + y_{SC}y_{aFCI} ) \nonumber\\
&\partial_l y_{aFCI} = (2-\Delta^{aFCI})y_{aFCI}
- (y_{FCI}y_{CDW} + y_{SC}y_{FCI} ) \nonumber\\
&\partial_l y_{SC} = (2-\Delta^{SC})y_{SC}
- y_{FCI} y_{aFCI} \nonumber\\
&\partial_l y_{CDW} = (2-\Delta^{CDW})y_{CDW}
- y_{FCI} y_{aFCI}
\ea 
where the scaling dimensions are given by \cref{eq:general_scaling_dimension}
\ba 
&\Delta_{FCI} =  \frac{9}{2}K_e + \frac{1}{2K_o} 
+ \frac{3}{4}\frac{-K_ev_1'+K_ov_2'}{K_o}
\nonumber\\
&\Delta_{aFCI} =   \frac{9}{2}K_e + \frac{1}{2K_o} 
- \frac{3}{4}\frac{-K_ev_1'+K_ov_2'}{K_o}
\nonumber\\
&\Delta_{CDW} = 18K_e 
\nonumber\\
&\Delta_{SC} = \frac{2}{K_o}
\ea 

We numerically solve the RG flow in \cref{eq:RG_flow_K,eq:RG_flow_y}, with initial conditions
\ba 
&K_e(l=0) = K_{e,0},\quad 
K_o(l=0) = K_{o,0}
,\quad v_1(l=0) =v_{2}(l=0) = 0 \nonumber\\
&y_{FCI}(l=0) = y_{FCI,0},\quad y_{aFCI}(l=0) = y_{aFCI,0} \nonumber\\
&y_{SC}(l=0) = y_{CDW}(l=0) =0 
\ea 
In practice, we assume that the initial microscopic values of $y_{CDW},y_{SC}$ are zero, and that nonzero $y_{CDW}$ and $y_{SC}$ are induced by fluctuations in the FCI and aFCI channels. 
We consider different combinations of $K_{e,0},K_{o,0}$ and take $y_{FCI,0}=0.1$ and $y_{aFCI,0} /y_{FCI,0} = 0.5$. The final phase of the system is determined by which coupling constant $y_\lambda$ first reaches $1$, indicating an instability of the SLL. 
The corresponding energy scale is determined by the critical value $l^*_\lambda$ defined by 
\ba 
y_\lambda(l^*_\lambda) = 1 \, . 
\ea 
The temperature scale of the system is then 
\ba 
T_{\lambda} \approx T_0e^{-l^*_\lambda }
\ea 
where $T_0$ denotes the initial UV energy scale of the system, which can be understood as the electronic bandwidth in the 1D limit. 

\begin{figure}
    \centering
    \includegraphics[width=1.0\linewidth]{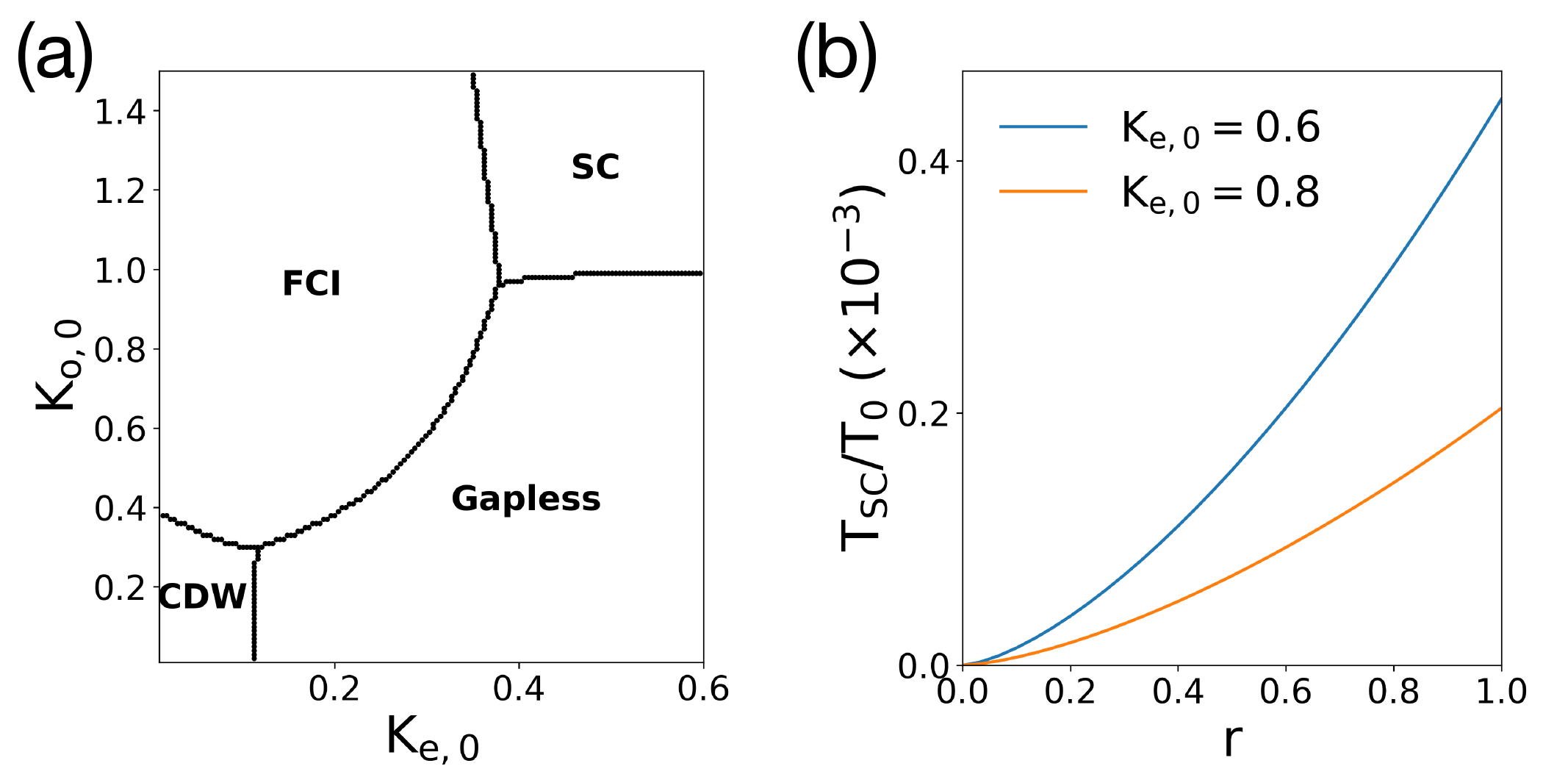}
    \caption{(a) Phase diagram obtained by solving the RG flow with initial conditions $K_{e}(l=0)=K_{e,0}$, $K_o(l=0)=K_{o,0}$, $y_{FCI}(l=0)=0.1$, $y_{aFCI}(l=0)=0.05$, and $y_{SC}(l=0)=y_{CDW}(l=0)=0$. 
    (b) Characteristic energy/temperature scale of the SC phase $T_{SC}$ as a function of $t_{x}'/t_x = y_{aFCI,0}/y_{FCI,0}$. We take parameters inside the SC phase with $K_o(l=0)=1.5$. Larger $t_x'/t_x$ indicates stronger quantum geometry, which in turn produces a more stable SC phase. $T_0$ denotes the UV energy scale of the system, proportional to the electronic bandwidth in the 1D limit. 
    }
    \label{fig:app_sc_process}
\end{figure}

\begin{figure}
    \centering
    \includegraphics[width=0.4\linewidth]{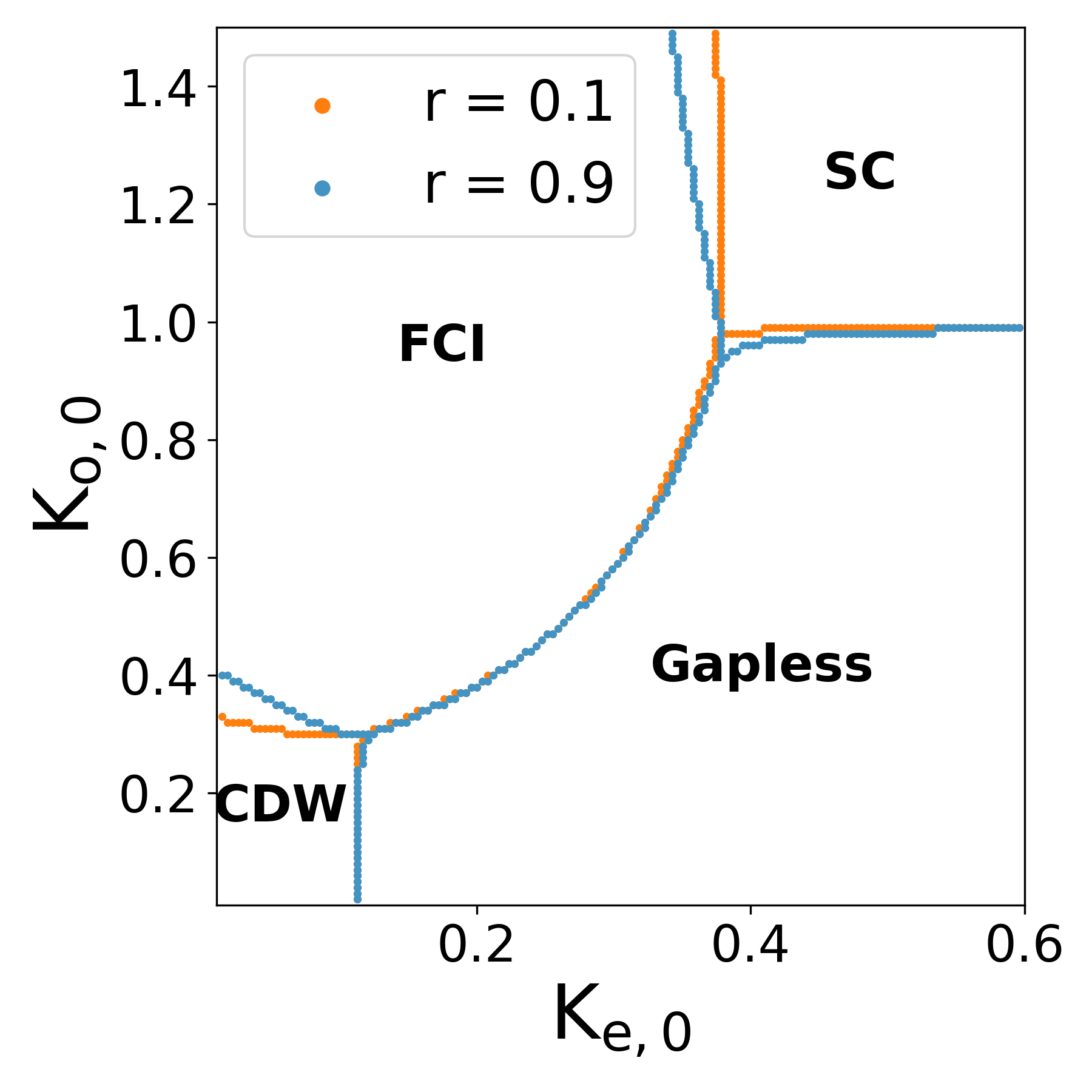}
    \caption{ Phase diagram at different $r$ obtained by solving the RG flow with initial conditions $K_{e}(l=0)=K_{e,0}$, $K_o(l=0)=K_{o,0}$, $y_{FCI}(l=0)=0.1$, $y_{aFCI}(l=0)=0.1r$, and $y_{SC}(l=0)=y_{CDW}(l=0)=0$.}
    \label{fig:phase_diagram_boundary_diff_r}
\end{figure}

The phase diagram obtained from the RG equations is shown in \cref{fig:app_sc_process}(a), where both SC and CDW phases appear near the FCI phase. 
We emphasize that the couplings $y_{CDW}$ and $y_{SC}$ naturally emerge during the RG flow, induced by the interplay between FCI and aFCI scattering processes. 
Although the location of the phase boundary is not significantly affected by variations in $r=t_x'/t_x$ (\cref{fig:phase_diagram_boundary_diff_r}), the transition temperature is sensitive to $r$ (\cref{fig:app_sc_process} (b)). 
Increasing $t_x'/t_x$ enhances the quantum geometry and makes the aFCI channel stronger relative to the FCI channel. This makes the SC phase more robust, as indicated by the increasing characteristic energy scale in \cref{fig:app_sc_process}(b).

\section{Filling $\nu=2/3$ and particle-hole transformation}
\label{app:nu_23}

We show that similar behavior, namely CDW and SC correlations induced by the cooperation of FCI and aFCI scattering processes, also appears at $\nu=2/3$.

We first consider the FCI phase at $\nu=2/3$, which can be obtained through a particle-hole transformation of the $\nu=1/3$ FCI phase. 
As mentioned in Ref.~\cite{PhysRevB.99.035130}, the particle-hole transformation within the wire construction is defined as 
\ba 
&\psi_{j,1}(r) 
\rightarrow \psi_{j,2}^\dag(r) \nonumber\\
&\psi_{j+1,2}(r) \rightarrow \psi_{j,1}^\dag(r) 
\ea 
Under the particle-hole transformation, the boson fields transform as
\ba 
&\theta_j(r) \rightarrow  \frac{1}{2}\bigg[-\theta_{j}(r) + \phi_j(r) -\theta_{j-1}(r) -\phi_{j-1}(r)\bigg] \nonumber\\
&\phi_j(r)\rightarrow \frac{1}{2}\bigg[-\theta_{j}(r) + \phi_j(r) +\theta_{j-1}(r) +\phi_{j-1}(r)\bigg] 
\ea 
The FCI term then transforms as 
\ba 
&\Theta_j^{FCI}(r) = \theta_j(r) -\theta_{j+1}(r) + 3(\phi_j(r) + \phi_{j+1}(r)) \rightarrow  \Theta_j^{\overline{FCI}}(r) =  4\phi_j(r) + \theta_{j-1}(r)+\phi_{j-1}(r) -\theta_{j+1}(r)+\phi_{j+1}  (r)
\ea 
As in the $\nu = 1/3 $ case, we can also introduce the aFCI term at $\nu=2/3$, which reads 
\ba 
\Theta_j^{\overline{aFCI}}(r) =  4\phi_j(r) - \theta_{j-1}(r)+\phi_{j-1} (r)+ \theta_{j+1}(r)+\phi_{j+1}(r)
\ea

As at $\nu=1/3$, the interplay between the $\overline{FCI}$ and $\overline{aFCI}$ channels also leads to SC and CDW correlations,
\ba 
&\Theta_j^{\overline{CDW}}(r) = \Theta_j^{\overline{FCI}}(r) + \Theta_j^{\overline{aFCI}}(r) = 8 \phi_{j} (r)+ 2\phi_{j-1} (r)+2\phi_{j+1} (r)\nonumber\\
&\Theta_j^{\overline{SC}}(r) =\Theta_j^{\overline{FCI}}(r) - \Theta_j^{\overline{aFCI}}(r) =2\theta_{j-1} (r)- 2 \theta_{j+1 }(r)
\ea 
This implies a similar OPE
\ba 
: e^{i\Theta_j^{\overline{FCI}}(R-\frac{r}{2})}:
:e^{i\Theta_j^{\overline{aFCI}}(R+\frac{r}{2})}: 
\sim 
\frac{1}{|r|^{\Delta^{\overline{FCI}}+\Delta^{\overline{aFCI}}-\Delta^{\overline{CDW}}}}e^{i\Theta_j^{\overline{CDW}}(R)} \nonumber\\ 
: e^{i\Theta_j^{\overline{FCI}}(R-\frac{r}{2})}:
:e^{-i\Theta_j^{\overline{aFCI}}(R+\frac{r}{2})}: 
\sim 
\frac{1}{|r|^{\Delta^{\overline{FCI}}+\Delta^{\overline{aFCI}}-\Delta^{\overline{SC}}}}e^{i\Theta_j^{\overline{SC}}(R)}
\ea  

In terms of the fermionic fields, 
\ba 
&e^{i\Theta_j^{\overline{CDW}}(r)} 
\sim [\psi_{j-1,1}^\dag(r) \psi_{j-1,2}(r) ]
[\psi_{j,1}^\dag(r) \psi_{j,2}(r) ]^2
[\psi_{j+1,1}^\dag(r) \psi_{j+1,2}(r) ] \nonumber\\
&e^{i\Theta_j^{\overline{SC}}(r)} 
\sim [\psi_{j-1,1}^\dag(r) \psi_{j-1,2}^\dag(r) ]
[\psi_{j+1,1}(r) \psi_{j+1,2}(r) ]
\ea 
Here $e^{i\Theta_{\overline{CDW}}}$ describes a coupling between particle-hole operators across three wires, while $e^{i\Theta_{\overline{SC}}}$ behaves as a Josephson coupling between wires and stabilizes an SC phase. 
In summary, we expect similar physics at filling $\nu=2/3$, where the SC and CDW instabilities emerge from the cooperation between FCI and aFCI scattering processes.

\end{document}